\documentclass[journal]{IEEEtran}

\ifCLASSINFOpdf
  \usepackage[pdftex]{graphicx}
\else
\fi

\usepackage{amsmath}
\usepackage{amssymb}
\usepackage{url}
\usepackage{comment}

\usepackage{amsmath,amsfonts,bm}

\def\eqref#1{equation~\ref{#1}}

\def\1{\bm{1}}

\DeclareMathAlphabet{\mathsfit}{\encodingdefault}{\sfdefault}{m}{sl}
\SetMathAlphabet{\mathsfit}{bold}{\encodingdefault}{\sfdefault}{bx}{n}

\usepackage[breaklinks=true,bookmarks=false]{hyperref}

\usepackage{microtype}      
\usepackage{xcolor}         
\usepackage{multirow}
\usepackage{graphicx}
\usepackage{amsmath}
\usepackage{caption}
\usepackage{subcaption}
\usepackage{booktabs}
\usepackage{enumitem}
\usepackage{xcolor}         

\begin{document}

\title{Scalable neural pushbroom architectures for real-time denoising of hyperspectral images onboard satellites}

\author{Ziyao~Yi, Davide Piccinini, Diego~Valsesia, \IEEEmembership{Member, IEEE}, Tiziano~Bianchi, \IEEEmembership{Member, IEEE}, and~Enrico~Magli, \IEEEmembership{Fellow, IEEE}
\thanks{The authors are with Politecnico di Torino -- Department of Electronics and Telecommunications, Italy. email: \{name.surname\}@polito.it. This study was carried out within the FAIR - Future Artificial Intelligence Research and received funding from the European Union Next-GenerationEU (PIANO NAZIONALE DI RIPRESA E RESILIENZA (PNRR) – MISSIONE 4 COMPONENTE 2, INVESTIMENTO 1.3 – D.D. 1555 11/10/2022, PE00000013, CIG B421A95680, CUP E13C22001800001). This manuscript reflects only the authors' views and opinions, neither the European Union nor the European Commission can be considered responsible for them.}
}

\maketitle

\begin{abstract}
The next generation of Earth observation satellites will seek to deploy intelligent models directly onboard the payload in order to minimize the latency incurred by the transmission and processing chain of the ground segment, for time-critical applications. Designing neural architectures for onboard execution, particularly for satellite-based hyperspectral imagers, poses novel challenges due to the unique constraints of this environment and imaging system that are largely unexplored by the traditional computer vision literature. In this paper, we show that this setting requires addressing three competing objectives, namely high-quality inference with low complexity, dynamic power scalability and fault tolerance. We focus on the problem of hyperspectral image denoising, which is a critical task to enable effective downstream inference, and highlights the constraints of the onboard processing scenario. We propose a neural network design that addresses the three aforementioned objectives with several novel contributions. In particular, we propose a mixture of denoisers that can be resilient to radiation-induced faults as well as allowing for time-varying power scaling. Moreover, each denoiser employs an innovative architecture where an image is processed line-by-line in a causal way, with a memory of past lines, in order to match the acquisition process of pushbroom hyperspectral sensors and greatly limit memory requirements. We show that the proposed architecture can run in real-time, i.e., process one line in the time it takes to acquire the next one, on low-power hardware and provide competitive denoising quality with respect to significantly more complex state-of-the-art models. We also show that the power scalability and fault tolerance objectives provide a design space with multiple tradeoffs between those properties and denoising quality. 
\end{abstract}

\section{Introduction}
\label{sec:intro}
\IEEEPARstart{H}{yperspectral} images (HSIs) \cite{qian2021hyperspectral} acquire a spatial-spectral signal across a wide range of the electromagnetic spectrum, typically tens or hundreds of wavelengths, thus capturing fine information about materials in the imaged scenes, which makes them widely used in multiple fields \cite{khan2018modern} such as environmental monitoring \cite{obermeier2019grassland}, urban planning \cite{weber2018hyperspectral}, and climate change \cite{durairaj2024sustainable}. 

Satellite HSIs are typically acquired by means of pushbroom sensors, which consist in a linear array of detectors in the across-track direction, i.e., the direction perpendicular to the satellite movement, for each spectral band. A pushbroom sensor thus acquires one line of an image with all its spectral bands within the sensor exposure time, and then the satellite movement in the along-track direction allows to capture successive lines. For instance, existing hyperspectral satellites such as PRISMA \cite{cogliati2021prisma}, EnMAP \cite{9484000} or HySIS \cite{eoportal_hysis} all use pushbroom sensors with largely similar characteristics. The line acquisition time is dictated by factors such as the desired ground sampling distance (typically 30m per pixel), the speed of the satellite due to its orbit as well as the achievable signal-to-noise ratio of the detectors, and it is around 4.5 ms for the aforementioned missions \cite{prismaATBD}. Limitations in the detector quality, as well as constraints on the exposure time, typically result in significant amounts of noise affecting the images, particularly at infrared wavelengths. Moreover, slightly different gains between detectors of adjacent pixels result in striping noise which may hinder image analysis. Therefore, in order to fully exploit the content of HSIs, denoising \cite{wei20203} is a critical preprocessing operation. 

At the same time, the space industry is undergoing a transformation, since it has recognized that the current system where images are downloaded to the ground segment within the limited windows of transmission and then processed with highly accurate but costly algorithms, leads to very long delays, potentially in the order of days, before products are available to the final users. The next generation of Earth observation satellites will seek to move as much processing onboard the satellites as possible to enable low-delay products such as alerts or detection results, which can be transmitted with priority due to their smaller sizes. However, this goal clashes with the severe constraints that onboard computing platforms face and calls for innovative designs that consider these constraints since their inception, rather than just scaling down models developed for more traditional scenarios. First and foremost, the available power on an Earth observation satellite is limited since it comes from solar panels, and it is also variable depending on the current orbital position and attitude changes \cite{lagunas2024performance} with respect to the Sun. Moreover, the effects of radiation needs to be considered for all onboard electronics. Traditionally, dedicated hardware with radiation-hardened designs was used, but, recently, there is a trend towards using commercial off-the-shelf components which allows to use the latest hardware advancements, and building radiation proofing systems around them, e.g., with external shielding and components redundancy where an operation may be performed in parallel by multiple systems and results cross-checked.

In this paper, we study how to design a neural architecture for real-time denoising of hyperspectral images onboard a satellite, so that the acquired images are effectively preprocessed for further downstream tasks executed onboard. We first present the unique design goals that this setting entails and then we introduce an approach which allows to control the tradeoffs between the competing objectives. We validate our approach by considering the HSI denoising task. This is a particularly interesting benchmark because careful denoising of HSIs is critical for any downstream inference task, including ones that could be executed onboard. Due to the high-dimensionality of HSI is also a good stress test for our proposed approach since i) HSIs may be complex to handle, especially in terms of memory requirements due to the large number of bands, and ii) the need for real-time operations to match the hyperspectral pushbroom acquisition and render the denoised images readily available for further usage.

More in detail, we discuss how the onboard setting requires to account for three competing design goals: i) high inference quality at low computational complexity; ii) power scalability; iii) fault tolerance. Concerning the first objective, we propose a novel denoiser architecture that is specifically tailored to match the acquisition process of pushbroom sensors. Such architecture works in a causal line-by-line fashion with a lightweight memory mechanism, based on Mamba \cite{fu2024ssumamba}, that allows to exploit the features of past lines to denoise the current one. This design greatly limits computational complexity and memory requirements, and can be tailored to run within the acquisition time of one line to enable real-time operations. This is contrast with conventional network designs which need to process large 2D image tiles, and suffer from high memory requirements and complexity.
Furthermore, the power scalability objective is concerned with the peculiarities of power availability in space. A power-scalable model allows to be run at multiple power levels, with consequently varying quality, to match time-varying power availability. As we shall see in the paper, better performance at lower power consumptions may come at the expense of top performance with the full power budget. This is an often neglected aspect in the literature for satellite onboard processing, resulting in suboptimal models that either need to be temporarily turned off or are oversimplified.
Finally, fault tolerance is required to make the model robust to radiation-induced errors in the space environment, such as corrupting weight values. We show that the latter two objectives can be tackled by designing a mixture of denoisers rather than a single static model, with a shared routing module that performs attention-based fusion and fault filtering. This allows to place individual denoisers in the mixture on independent accelerators which can be smaller and easier to manufacture, can also be turned off to match the time-varying power requirements, and provide a sort of redundancy to faults. 

Our main contributions may thus be summarized as follows:
\begin{itemize}
    \item We introduce a set of design goals for neural architectures intended for onboard satellite deployment, explicitly addressing real-time constraints, power scalability, memory limits, and radiation-induced faults.
    
    \item We develop a line-wise hyperspectral denoising architecture and integrate it into a mixture-of-denoisers framework that together enable real-time processing aligned with pushbroom acquisition, while providing both power scalability and intrinsic fault tolerance.

    \item We propose training and fault-filtering strategies to support flexible power scalability tradeoffs and robustness under radiation-induced model degradation.
\end{itemize}

\section{Background and Related Work}
\label{sec:bkg}
\subsection{HSI Denoising}
HSI denoising is an ill-posed inverse problem, that has a long history of methods attempting the reconstruction of a clean image from noisy observations. Mathematically, the observed image $\mathbf{y}$ is modeled as the summation between noise $\mathbf{\epsilon}$ and the original image $\mathbf{x}$, i.e., $\mathbf{y} = \mathbf{x} +\mathbf{\epsilon}$.

Classical model-based approaches to HSI denoising leverage handcrafted priors such as sparse representations \cite{zhuang2018fast}, low-rank representations \cite{zhang2013hyperspectral}, and total variation (TV) minimization \cite{7438863}. BM3D \cite{4271520} and BM4D \cite{6253256} introduce the concept of grouping and collaborative filtering to exploit both local and non-local correlations. 
LRTFL0 \cite{xiong2019hyperspectral} restores HSI by approximating them via low-rank block term decomposition. Takemoto et al. \cite{takemoto2022graph} propose Graph-SSTV, a TV-based regularization method that constructs a graph capturing the spatial structure of the target HSI. 

In terms of learning-based methods, T3SC \cite{bodrito2021trainable} designs an architecture that is an unrolled version of a sparse coding method. 
MAC-Net \cite{9631264} combines model-based spectral low-rank structures and a deep spatial prior. Pang et al. \cite{pang2022trq3dnet} propose a two-branch architecture to extract the spectral correlation between different bands and individually exploit the global and local spatial features.  SERT \cite{li2023spectral} presents a spectral enhanced rectangle Transformer, driving it to explore the non-local spatial similarity and global spectral low-rank property of HSIs pixels. SSRT \cite{fu2024hyperspectral} combines global spectral correlation and non-local spatial self similarity properties within a single SSRT block. Moreover, state space models (SSMs) have recently been adopted for HSI tasks due to their ability to model long-range dependencies efficiently \cite{fu2024ssumamba}. 
Despite their promising performance, these methods remain impractical for onboard deployment in spacecraft, primarily due to constraints on computational resources, which makes real-time processing infeasible. In particular, they all have significant memory requirements as well as large computational complexities in terms of FLOPs per pixel. Moreover, none of them matches the acquisition of pushbroom sensors, causing the need to buffer a large number of lines for their operation. 
A few works adopt mixtures of models for image restoration tasks, typically to have multiple experts for different subtasks.  BDE \cite{liu2022robust} fuses pretrained denoisers in a pixel‑wise manner to handle heterogeneous, real‑world noise. Jiang \cite{8656554} combines several models to improve performance in Single Image Super-Resolution. EnsIR\cite{sun2024ensir} casts ensemble learning with a Gaussian mixture model (GMM) to better predict the weight of each single model.


\subsection{Deep sequence modeling} 
A classical approach to sequence modeling is based on Causal Convolutions, that confine each filter’s receptive field to previous timesteps only, preserving the autoregressive order \cite{van2016pixel}. Alternatively, Recurrent Neural Networks (RNNs) operate by recursively updating a hidden state as new inputs arrive. A key limitation of RNNs arises when dealing with long sequences, where the models may suffer from vanishing or exploding gradients. Long Short-Term Memory (LSTM) networks \cite{hochreiter1997long}, and Gated Recurrent Units (GRUs) \cite{chung2014empirical} introduced gating mechanisms to address this issue, but they can still face difficulties in modeling very long-range dependencies.

Transformers \cite{vaswani2017attention} rely on attention mechanisms, instead of recurrence, which enable input-dependent operations that capture long-range interactions across the sequence. 
They have become the 
standard for sequence processing in many domains, such as for text \cite{touvron2023llamaopenefficientfoundation}, image \cite{hatamizadeh2024diffit} and video generation \cite{brooks2024video} .
However, the quadratic computational and memory complexity of the attention with respect to sequence length remains a major bottleneck.
Various techniques have been proposed to mitigate this, including sparse attention \cite{child2019generating} and linear attention \cite{katharopoulos2020transformers}, but the scalability challenges persist.

SSMs have recently gained popularity due to their linear complexity with the sequence length. 
They model a sequence-to-sequence mapping that transforms an input sequence of features into an output sequence by an implicit internal latent state and that evolves through linear dynamics specified by system matrices. Gu et al. \cite{gu2021efficiently} demonstrated that carefully constraining the SSM parameters substantially improves model performance in practice. 
Extending this line of research 
\cite{fuhungry,peng2023rwkv,poli2023hyena}, Gu et al. \cite{gu2023mamba} further introduced Selective SSMs, an enhanced approach where SSM parameters dynamically depend on input data, allowing for a selectivity mechanism that effectively determines when and how information should be propagated or discarded throughout a sequence. Mamba has already shown considerable results across several domains, including audio and language, and has been extended into the area of computer vision\cite{zhuvision, liu2024vmamba}.
In the field of hyperspectral imagery, Mamba has also been deployed with very promising results \cite{li2024mambahsi, yao2024spectralmamba, sheng2024dualmamba, huang2024spectral}.
Nonetheless, the deployment of Mamba Blocks for denoising images in an incremental, line-by-line fashion remains unexplored. 

\begin{figure*}[t]
    \centering
    \includegraphics[width=0.9\textwidth]{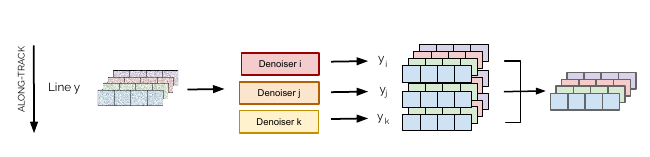}
    \caption{The acquired line gets processed by each independent line-based denoiser, which makes use of a compact memory of past lines to refine it. The different denoised lines are then aggregated to form the output.}
    \label{fig:overview}
\end{figure*}

\subsection{Onboard processing} 
In recent years, onboard processing has gained significant attention as an effective strategy for performing real-time satellite data analysis. 
By processing imagery directly onboard the satellite rather than transmitting extensive raw datasets to Earth, satellites can achieve more agile operational responsiveness, rapidly detect critical phenomena, and enable autonomous decision-making capabilities. 
Yao et al. \cite{yao2019board} introduced lightweight deep learning models on small satellites for maritime vessel detection. Giuffrida et al. \cite{giuffrida2021varphi} demonstrated the use of CNNs on the Phi-Sat-1 mission for real-time cloud detection. Building upon these foundational studies, Ziaja et al. \cite{ziaja2021benchmarking} provided comprehensive benchmarks of neural network architectures tailored for space-grade computing devices, and \cite{ruuvzivcka2022ravaen} proposed a streamlined variational autoencoder for efficient onboard change detection. Moreover, in \cite{inzerillo2024efficient}, the authors presented a modular framework specifically designed for multitask inference onboard satellites. 
Recently, Valsesia et al. \cite{valsesia2024onboard} proposed onboard compression of hyperspectral images with a predictive coding mechanism that predicts the next line using RWKV \cite{peng2023rwkv}, while Piccinini et al. \cite{p1, piccinini2025onboard} introduced an onboard super-resolution strategy that frames the problem as a previous low resolution line interpolation one.
Existing works do not paint a complete picture of onboard processing as they mostly focus on scaled-down designs to limit complexity but neither they consider power scaling or fault resilience goals nor they match the line-wise nature of pushbroom acquisiton. 


\section{Method}
\label{sec:method}

\begin{figure*}[t]
  \begin{minipage}[c]{0.48\textwidth}
    \vspace*{\fill}
    \centering
    \subcaptionbox{Overall architecture\label{fig:ensemble}}[\textwidth]{
      \includegraphics[width=\linewidth]{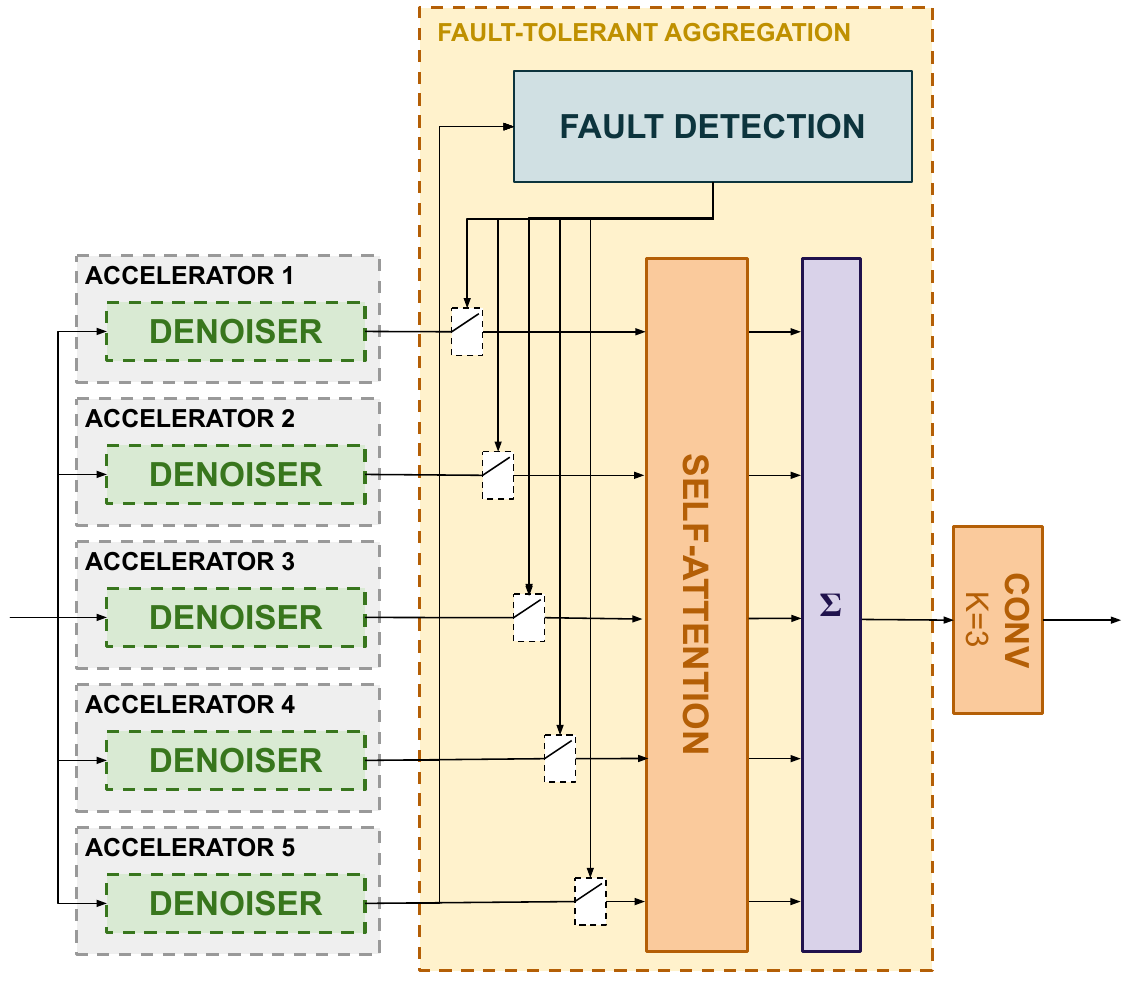}
    }
    \vspace*{\fill}
  \end{minipage}
  \hfill
  \begin{minipage}{0.48\textwidth}
    \subcaptionbox{Denoiser Architecture\label{fig:unet}}[\textwidth]{
      \includegraphics[width=\linewidth]{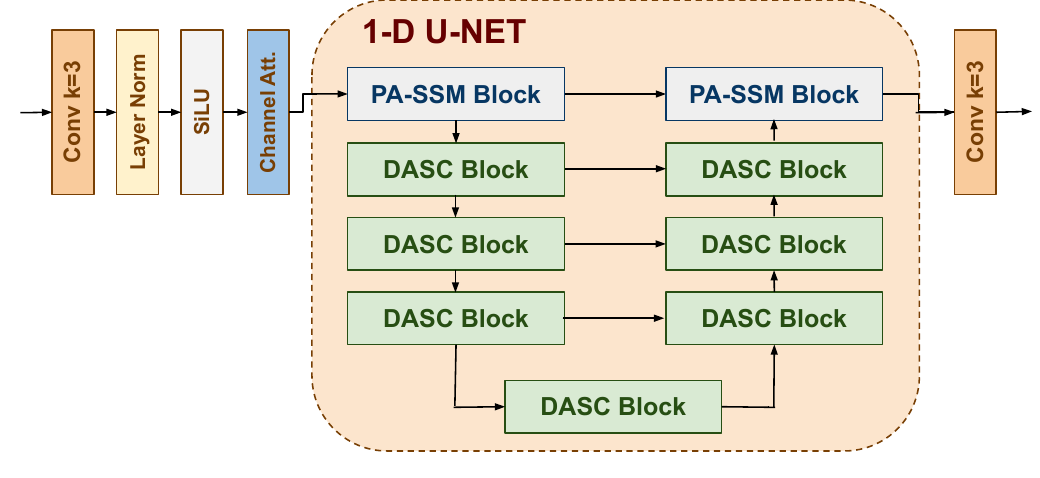}
    }
    
    \vspace{0.6ex} 
    
    \subcaptionbox{PA-SSM Block\label{fig:nafmamba}}[\textwidth]{
      \includegraphics[width=\linewidth]{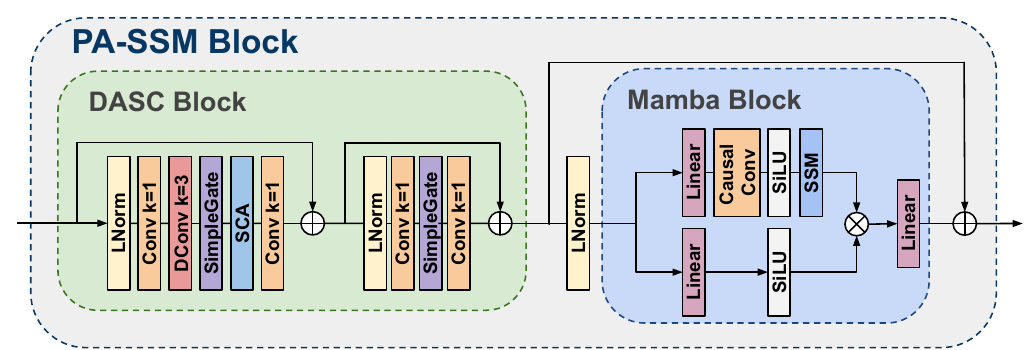}
    }
  \end{minipage}
  \vspace{-5pt}
  \caption{Proposed design. a) overall architecture with a mixture of $D$ denoisers running in parallel: each denoiser is placed on an independent hardware accelerator that can be turned on/off for power scaling; the output features are then checked for faults and aggregated with an attention mechanism. b) individual denoiser architecture, which is a 1-D UNet processing a line with a memory of past lines, consisting of 1-D DASC Blocks \cite{chen2022simple} and 1-D PA-SSM Blocks (c)).}
  \label{fig:layout}
  \vspace{-15pt}
\end{figure*}

This section presents the main contributions of our work. We start with a formal introduction of the goals that one must consider when designing neural architectures for onboard HSI processing, and we focus on the scenario of a satellite with a hypespectral pushbroom sensor. In the following subsections, we present the ingredients of the proposed design matching those goals and highlight how their competing nature leads to multiple tradeoffs.
We thus define the following design goals:

\begin{description}[style=nextline, labelwidth=0cm, leftmargin=!, labelsep=1em]
    \item[\textbf{Goal 1 [Low-complexity high-quality inference]}] 
The model should be designed to be run in real-time on low-power hardware matching the pushbroom acquisition dynamic. The quality of the output should be as competitive with the state-of-the-art as possible given the complexity constraint. 
    \item[\textbf{Goal 2 [Power scalability]}] 
    The model should be able to run at different power levels, allowing switching to a lower or higher level during an orbit.
    \item[\textbf{Goal 3 [Fault tolerance]}]
    The model should be designed to be resilient to radiation-induced faults, detecting them  when they happen and providing degraded outputs rather than breaking down.
\end{description}

The first goal guides the design to maximize the effectiveness of the neural model under a complexity constraint. For HSI processing, it is important that the model limits memory consumption, which could otherwise be significant due to the large number of spectral bands, and that the amount of operations to be performed is compatible with the line acquisition time, for real-time performance.

Power scalability suggests that during an orbit (or the satellite lifetime) different power constraints are available. The objective seeks to maximize the effectiveness of the model by producing an output, albeit degraded, even when a lower power is available. Notice that the specific amount of time spent at different power levels may guide the design to prioritize one over the other (e.g., if the maximum power is rarely available, design could prioritize the performance of the model at lower power levels).

Fault tolerance requires the design to take into account radiation-induced faults, according to some fault model. For example, according to \cite{cai2024evaluation}, each neural network weight sitting in long-term memory may be corrupted. The corruption is usually represented with a large deviation from the nominal value, as it would happen in case of flips in the most significant bits. A fault-tolerant model should be able to withstand some level of faults with a slightly degraded output rather than breaking down.

\subsection{Neural Network Architecture}

In this section we present the proposed neural network architecture for onboard HSI denoising that satisfies the previously outlined design goals. The overall workflow is shown in Fig. \ref{fig:overview}: the acquired line is processed by a mixture of independent denoisers, the outputs are dynamically aggregated and averaged. A high-level overview of the architecture is presented in Fig. \ref{fig:ensemble}. In the remainder of the paper, we denote by $N_l$, $N_c$, $N_b$ the number of along-track lines, across-track columns and spectral bands of each image, respectively.

The architecture is designed as a mixture of $D$ lightweight denoisers operating in parallel. The denoisers share their architecture, but they have independent parameters, i.e., they are independently trained from different random initializations to ensure a complementary ensemble. The feature maps produced by each not-faulty denoiser are aggregated by means of a residual self-attention operation and averaged before being projected to the image space. As common practice in denoising architectures, this constitutes as an estimate of the noise, which is then subtracted from the noisy input to output the denoised image. More formally, let $\mathbf{y}$ be the noisy image, and let $\mathbf{h}^{(d)} \in \mathbb{R}^{1 \times N_c \times F}$ be the feature maps output by denoiser $d$, then the aggregation stage computes:
\begin{align}
    \hat{\mathbf{x}} &= \mathbf{y} - \hat{\mathbf{n}} \\
    \hat{\mathbf{n}} &= \text{conv}\left( \frac{1}{|\mathcal{D}|} \sum_{d\in\mathcal{D}} \mathbf{z}^{(d)} \right)\\
    \mathbf{z}^{(d)} &= \text{SA}(\lbrace \mathbf{h}^{(i)} \rbrace_{i \in \mathcal{D}})^{(d)} + \mathbf{h}^{(d)} \label{eq:aggr} \\
    \begin{split}
            \mathcal{D} &= \mathcal{F}(\lbrace \mathbf{h}^{(i)} \rbrace_{i=1}^D) \\
    &= \left\lbrace i \in \lbrace 1,2,\dots,D \rbrace \text{ s.t. $i$ is not faulty} \right\rbrace 
    \end{split}
\label{eq:fault_detection}
\end{align}
where $\mathcal{F}$ is the fault detection method described in Sec. \ref{sec:fault}, which determines if a denoiser is affected by a fault and returns the set of working denoisers $\mathcal{D}$. The self-attention (SA) operation is the scaled dot-product attention \cite{vaswani2017attention}, operating on the sequence of $|\mathcal{D}|$ not-faulty denoisers, for each pixel.

The reason for a mixture design can be traced to Goals 2 and 3, which suggest a design not relying on a single large model. In particular, each individual denoiser can be placed on a small independent accelerator, so that each can be turned on or off to meet the power requirement. A training process to tune the desired tradeoff between performance at low power level or performance at high power levels can be devised (see Sec. \ref{sec:power}).
Moreover, a mixture design is inherently more resilient to faults since a fault in one denoiser will not affect the others, if the denoiser can be reliably filtered from contributing to the output. Indeed, one may think of this mixture design as being inspired by the traditional approach used in space of redundant computations, where multiple computers run the same computation in parallel and check each other's output. However, in this case there is no repeated calculation as each denoiser has its own weights and running more than one improves the final result.

The design of each denoiser is guided by Goal 1 in that it needs to be lightweight and match the pushbroom acquisition of the image. The primary challenge is in exploiting both local and long-range spatial context, while respecting stringent constraints on memory footprint and computational cost. For this reason, we propose a novel lightweight architecture that processes an image line-by-line, as it is acquired by the sensor. More in detail, the currently-acquired line with all its spectral bands is processed to extract joint across-track/spectral features. However, the model also needs a memory of past lines in order to exploit inter-line, possibly long-range, correlations among such features. For this purpose we exploit Mamba \cite{gu2023mamba}, a recently-proposed SSM, which allows to efficiently capture a memory of the features of past lines in a compact state. As mentioned in Sec. \ref{sec:bkg}, Mamba offers a great tradeoff between complexity and performance for this scenario compared to other sequence processing options such as transformers for which KV caching would require large amounts of memory, outperforming RNNs and simple causal convolution. An overview of the denoiser architecture is presented in Fig.~\ref{fig:unet}.

Since the model processes one line at a time, all the operations are spatially 1-D, with the along-track dimension fixed to $1$. 
First, the $N_b$ spectral channels are projected into a more compact feature space of size $F$ via a learnable 1-D convolution, which also allows to manage memory consumption. In fact, since redundancy in the spectral dimension is generally very high, a modest number of features is generally suitable to properly represent the original bands.
This yields shallow features for each row, which are then refined by LayerNorm, followed by a non-linearity (SiLU) and a traditional channel attention block \cite{woo2018cbam}. 
These features from the current line are then processed by a 1-D U-Net, which largely consists in blocks (DASC Block) inspired by the NAFNet architecture for image restoration \cite{chen2022simple}, adapted to 1-D processing. Each DASC Block provides a lightweight attention-like mechanism to extract across-track/spectral features via the SimpleGate and simplified channel attention operations. We refer the reader to \cite{chen2022simple} for details of such layers. Downsampling is implemented using a stride-$2$ 1-D convolution, whereas the upsampling makes use of a stride-$2$ 1-D transpose convolution. Skip connections are implemented with feature addition.

The first and last blocks of the U-Net, i.e., the ones operating at the full spatial resolution, are extended with a residual Mamba Block. The Mamba Block has an internal arbitrary feature expansion factor $E$, while the last linear layer of the block returns to the original feature size $F$. Mamba leverages a causal convolution with a Selective State-Space module to model the inter-line context with minimal state storage. This operation is performed in parallel for each feature vector representing a pixel in the current line.  More formally, let $l$ be the index of the currently processed line. Then we have  
 \begin{align}
     \mathbf{v'}_{l} &= \mathrm{CausalConv} \bigl( \mathbf{v}_{l} \bigr) \in \mathbb{R}^{N_c \times (FE) \times 1},
 \end{align}
where the operation is a 1-D causal convolution with kernel size $K$ 
that leverages an internal state comprised of a memory of $K$ 
lines of features, i.e. the $K-1$ 
previous lines 
and the current one, to produce tensor $\mathbf{v'_{i}}$. 
After a SiLU nonlinearity, the features are fed to the SSM module,  where they undergo the following operations:       
 \begin{align}
\mathbf{h}_{l}(t) &= \mathbf{A}\mathbf{h}_{l-1}(t) + \mathbf{B}\mathbf{v'}_{l}(t),\\
\mathbf{v''}_{l}(t) &= \mathbf{C}\mathbf{h}_{l}(t) + \mathbf{D}\mathbf{v'}_{l}(t),
\end{align}
where $\mathbf{h}\in \mathbb{R}^{N_c \times (FE)\times D}$  is the internal state of the model and carries on the knowledge of all the past lines. The dimension $D$ is called the state factor, and controls how big the internal state has to be to effectively store and process all past and current information. 
Notice that the SSM works in parallel for each pixel in the line, leading to the internal state tensor $\mathbf{h}$ having an independent memory for each across-track pixel. Thanks to the properties of the Mamba Block, the model can effectively store and retrieve information from the past lines.

\subsection{Power Scalability}
\label{sec:power}
The power scalability goal we set out for onboard design postulates that our proposed model needs to be able to run at different power levels. This is achieved by designing a $D$-denoiser mixture model using $D$ different hardware accelerators. Each of these accelerators can be turned on or off as desired, thus some are disabled for times with lower power budgets. 

Since the power budget is typically variable over time, we now show how knowledge of its distribution can be used to guide the training process of the mixture model to different tradeoffs. In particular, we show that training can optimize the mixture to use a smaller number of denoisers, at the expense of the peak performance when more denoisers are used, or viceversa. In other words, the training process can skew the performance of the mixture towards subsets of denoisers that are regarded as a more important scenario in the power budget: a scenario in which it is rare that all denoisers are used, but most of the time 2 or 3 are, will seek to maximize the performance of the mixture with the latter number of denoisers, at the expense of performance with all denoisers. 
To achieve this, we propose a denoiser sampling scheme for training, which randomly samples only a subset of denoisers to be used for a minibatch. A user-defined probability distribution is designed to determine the probability of using $N \leq D$ denoisers in each training minibatch. Specifically, we design this distribution as a discrete exponential controlled by a user-defined parameter $\lambda$. Let us call $s$ the number of sampled denoisers, then the probability of using $1\leq N \leq D$ denoisers for the current training iteration is:
\begin{align}
    \mathbb{P}\left( s = N \right) = \frac{e^{\lambda N}}{\sum_{j=1}^{D} e^{\lambda N}}
\end{align}
We call $\lambda$ the power-scalability factor which can be setup by the system designer according to the desired power scalability properties. Notice that when $\lambda=0$ all subset cardinalities are equally likely, meaning that there is no preferred power level. When $\lambda$ assumes large positive values, training will be more likely to always choose to use all denoisers thus the system will optimize for the maximum power level. Conversely, large negative values will optimize the system to use one or few denoisers at the expense of performance when all denoisers are used.



\subsection{Fault Tolerance}
\label{sec:fault}

As anticipated in Eq. \ref{eq:fault_detection}, the aggregation function of the mixture model can discard denoisers deemed faulty in order to be resilient to radiation-induced errors. We thus present a detection method that implements function $\mathcal{F}$ in the aforementioned equation.
We assume a fairly common fault model in which a weight of a denoiser is perturbed by a large deviation, emulating a bit flip in a significant bit. For simplicity, we also assume that the probability of a fault happening in the aggregation and fault filtering steps is small compared to a fault in one or more denoisers.    
We propose a detection method that relies on the key, and query projections of the pixel features that are used for aggregation by means of self-attention. In particular, we look at the diagonal of the attention score matrix $\text{softmax}\left(\mathbf{k}\mathbf{q}^T / \sqrt{F} \right)$, i.e., the scalar product between the key and query projections of each denoiser (after row softmax normalization). Since an attention matrix is generated for each pixel, we look for spatial variance of each value on the diagonal as a metric of reliability. A low spatial variance should represent the nominal case in which the SA aggregation weight for the current denoiser does not vary wildly over the space dimension. A high variance instead is indicative of a fault as it creates inconsistent features.
Hence, the proposed implementation of fault detection is:
\begin{align}
   j \in \mathcal{D} &= \mathcal{F}(\lbrace \mathbf{h}^{(i)} \rbrace_{i=1}^D) \quad \text{iff} \quad  \mathbb{V}ar\left[ A_{jj} \right] > \tau \\
    A &= \text{softmax}\left(\mathbf{k}\mathbf{q}^T / \sqrt{F} \right), \quad \\ \nonumber
   \mathbf{k}&=[\mathbf{h}^{(1)};\dots;\mathbf{h}^{(D)}] W_k, \quad \\ \nonumber
   \mathbf{q}&=[\mathbf{h}^{(1)};\dots;\mathbf{h}^{(D)}] W_q
\end{align}
Notice that while the full $D \times D$ attention matrix is computed for fault detection, only the $|\mathcal{D}|$ not-faulty denoisers are used in the aggregation with a reduced $|\mathcal{D}| \times |\mathcal{D}|$ attention matrix (see Eq. \ref{eq:aggr}). The threshold $\tau$ can be cross-validated at training time to optimize the accuracy of detection process or other metrics such as precision or recall.

\section{Experimental Results}
\label{sec:experimental}

\begin{figure}
    \centering
    {\includegraphics[width=\linewidth]{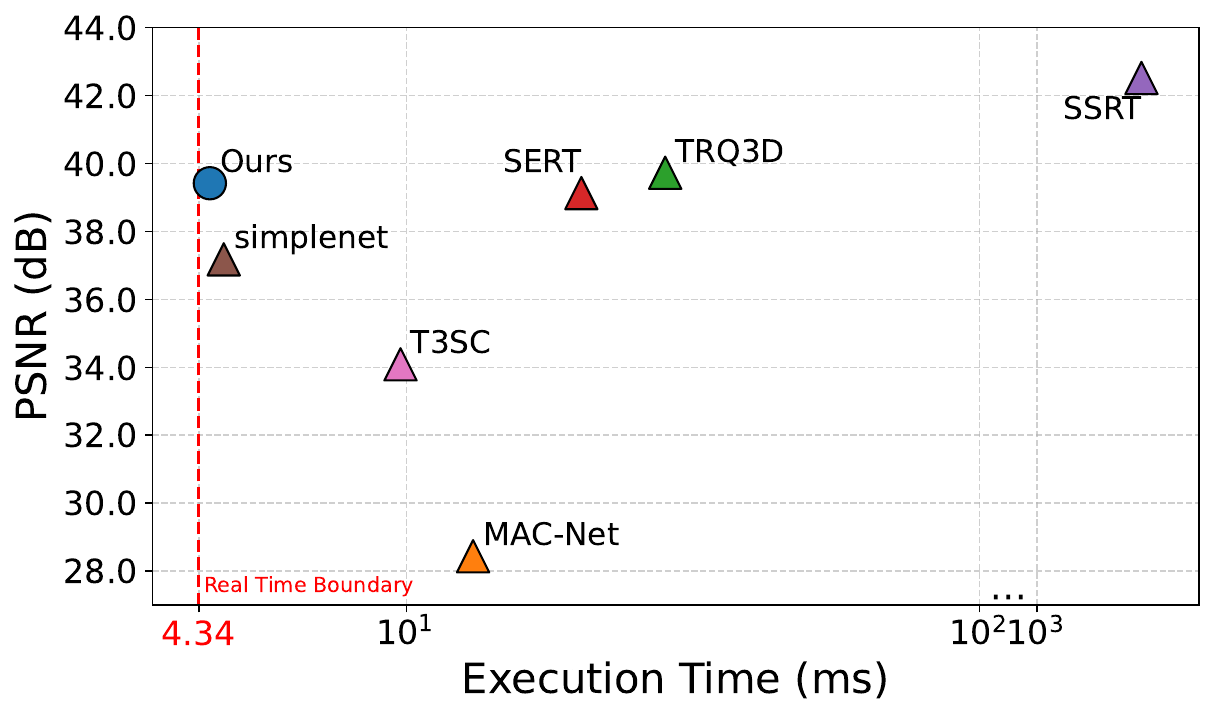}}
    \caption{Runtime on Nvidia Jetson Orin Nano normalized for a $1 \times 1000 \times 66$ line. 4.34ms is the Line Acquisition Time of the PRISMA satellite.}
    \label{fig:runtime}

  \vspace{-5pt}
\end{figure}

 \vspace{-5pt}

\begin{figure*}[htbp]
\centering
 \captionsetup{font=small}
 \captionsetup[sub]{font=scriptsize,skip=2pt} %
\captionsetup[subfigure]{font=scriptsize}
\begin{subfigure}[b]{0.124\linewidth}
  \includegraphics[width=\linewidth]{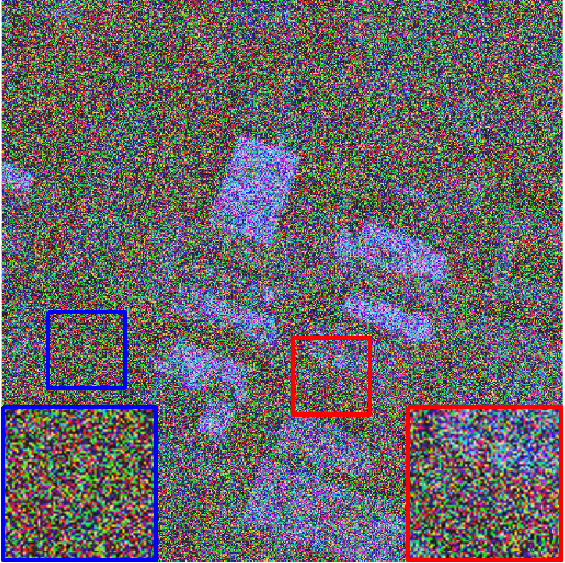}
  \caption{Noisy}
  \label{fig:houston_noisy}
\end{subfigure}
\begin{subfigure}[b]{0.124\linewidth}
  \includegraphics[width=\linewidth]{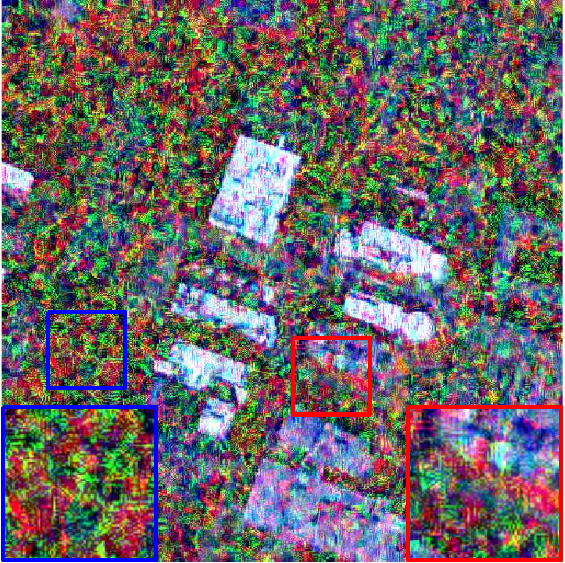}
  \caption{BM4D}
  \label{fig:houston_BM4D}
\end{subfigure}
\begin{subfigure}[b]{0.124\linewidth}
  \includegraphics[width=\linewidth]{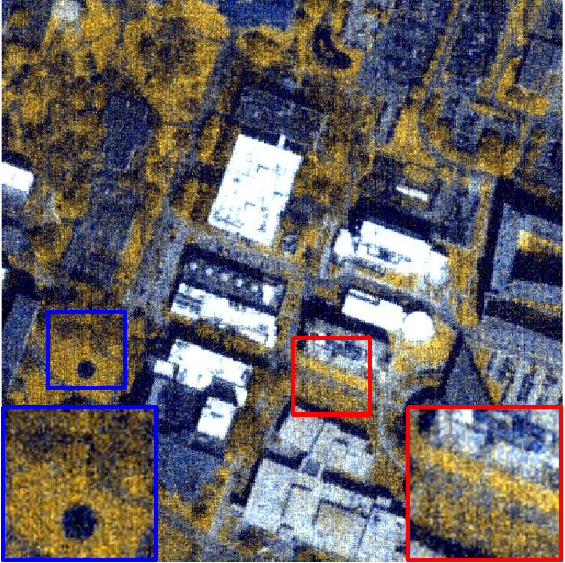}
  \caption{$LRTFL_0$}
  \label{fig:houston_lrtfl0}
\end{subfigure}
\begin{subfigure}[b]{0.124\linewidth}
  \includegraphics[width=\linewidth]{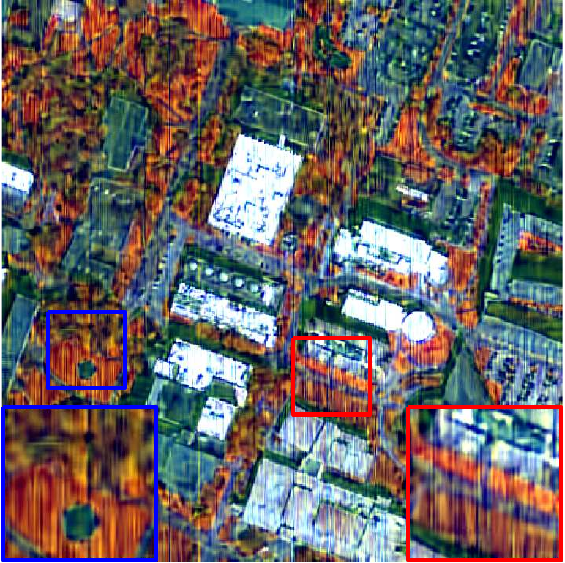}
  \caption{FastHyDe}
  \label{fig:houston_FastHYDe}
\end{subfigure}
\begin{subfigure}[b]{0.124\linewidth}
\includegraphics[width=\linewidth]{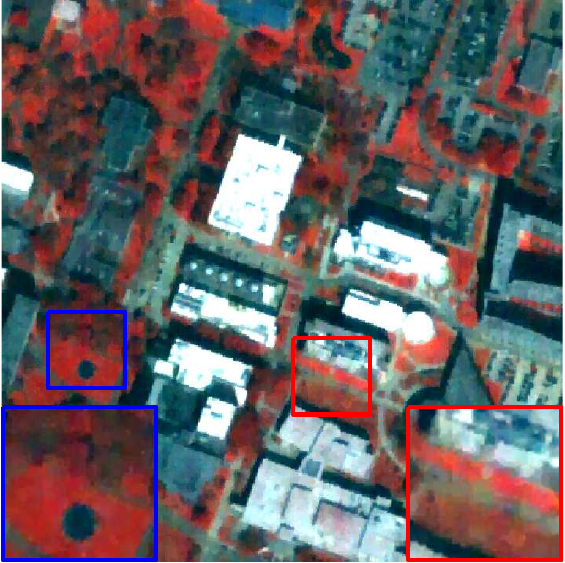}
  \caption{E3DTV}
  \label{fig:houston_e3dtv}
\end{subfigure}
\begin{subfigure}[b]{0.124\linewidth}
\includegraphics[width=\linewidth]{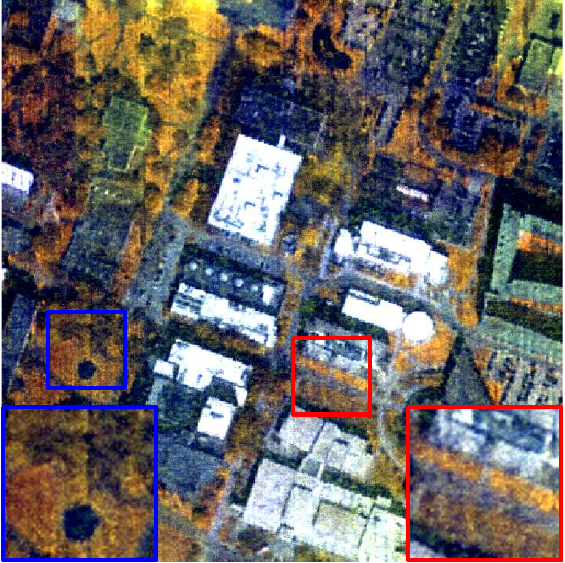}
  \caption{MAC-Net}
  \label{fig:houston_MAC-Net}
\end{subfigure}
\begin{subfigure}[b]{0.124\linewidth}
\includegraphics[width=\linewidth]{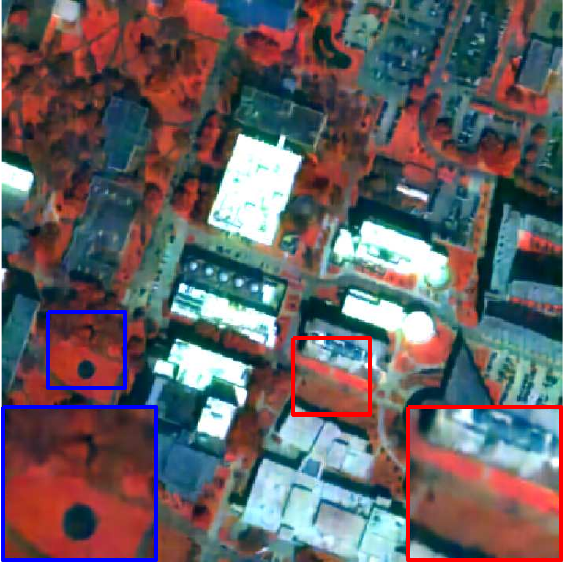}
  \caption{T3SC}
  \label{fig:houston_t3sc}
\end{subfigure}
\begin{subfigure}[t]{0.124\linewidth}
\includegraphics[width=\linewidth]{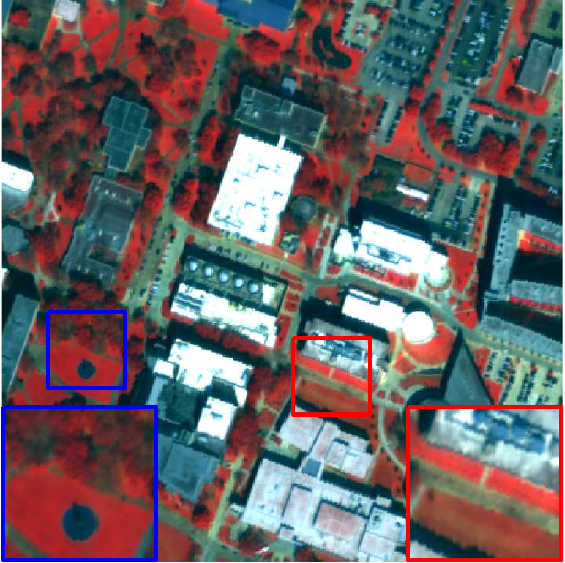}
  \caption{Clean\\ }
  \label{fig:houston_clean}
\end{subfigure}
\begin{subfigure}[t]{0.124\linewidth}
\includegraphics[width=\linewidth]{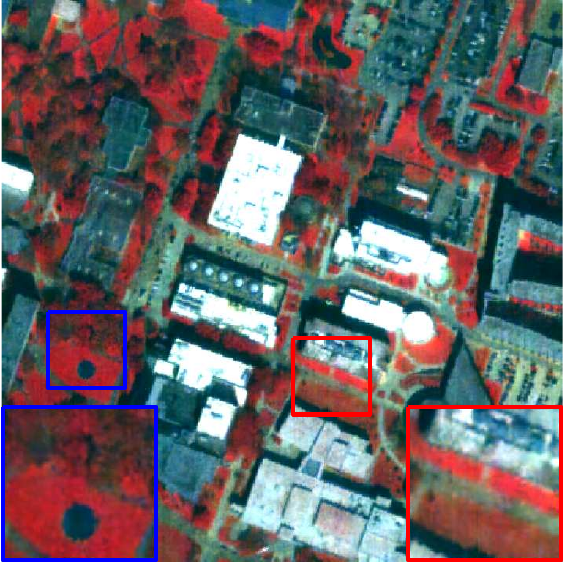}
  \caption{TRQ3D\\}
  \label{fig:houston_trq3d}
\end{subfigure}
\begin{subfigure}[t]{0.124\linewidth}
\includegraphics[width=\linewidth]{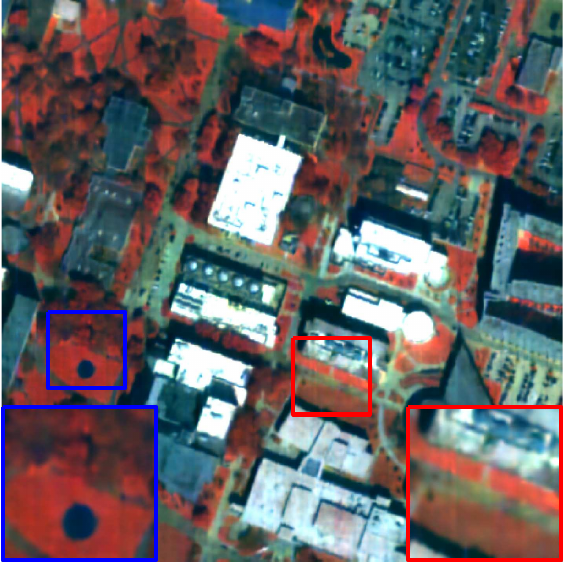}
  \caption{SERT\\}
  \label{fig:houston_sert}
\end{subfigure}
\begin{subfigure}[t]{0.124\linewidth}
\includegraphics[width=\linewidth]{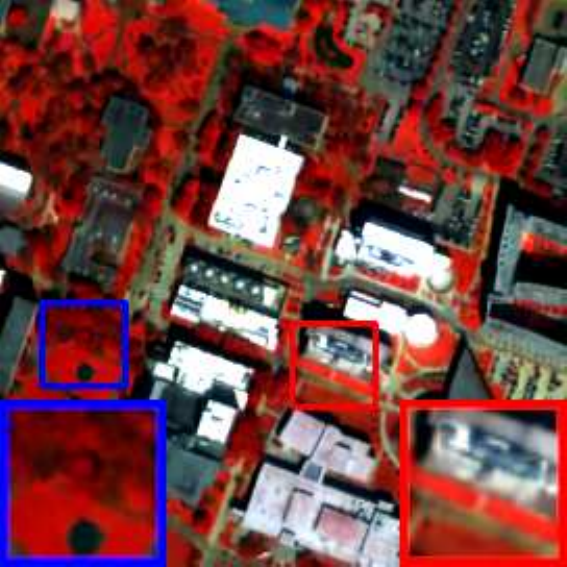}
  \caption{SSRT\\}
  \label{fig:houston_SSRT}
\end{subfigure}
\begin{subfigure}[t]{0.124\linewidth}
\includegraphics[width=\linewidth]{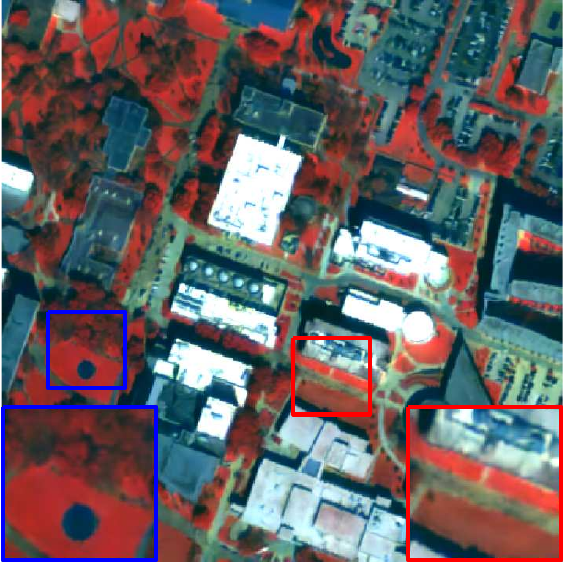}
  \caption{SSUMamba\\}
  \label{fig:houston_mamba}
\end{subfigure}
\begin{subfigure}[t]{0.122\linewidth}
  \includegraphics[width=\linewidth]{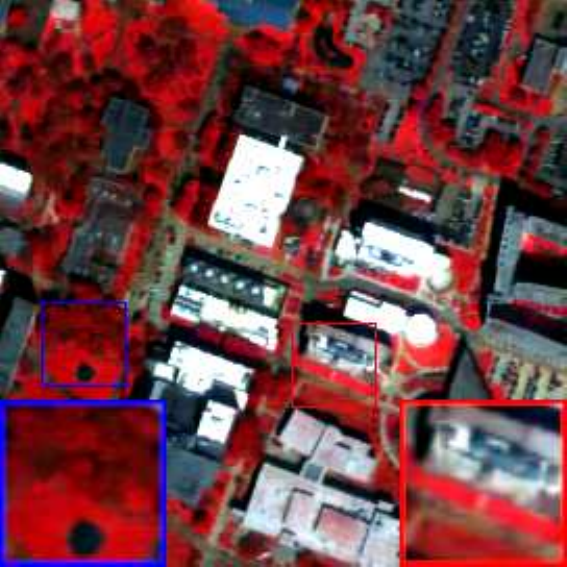}
  \caption{Proposed\\(5-mixture)}
  \label{fig:houstun_5-mixture}
\end{subfigure}
\begin{subfigure}[t]{0.122\linewidth}
  \includegraphics[width=\linewidth]{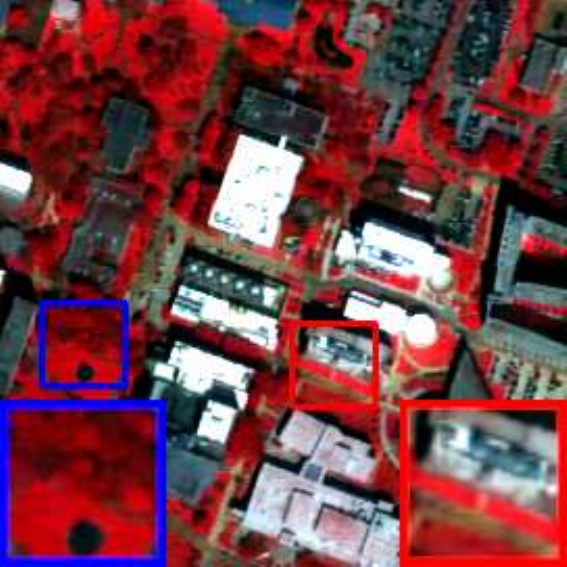}
  \caption{Proposed\\(1-large)}
  \label{fig:houston_1large}
\end{subfigure}

\caption{Denoising results on the Houston 2018 HSI with mixture noise. The false-color images are generated by combining bands 35th, 20th, and 5th.}
\label{fig:houston_visual}
\end{figure*}

\begin{figure*}[t]
\centering
 \captionsetup{font=small}
 \captionsetup[sub]{font=scriptsize,skip=2pt} %
\captionsetup[subfigure]{font=scriptsize}
\begin{subfigure}[b]{0.124\linewidth}
  \includegraphics[width=\linewidth]{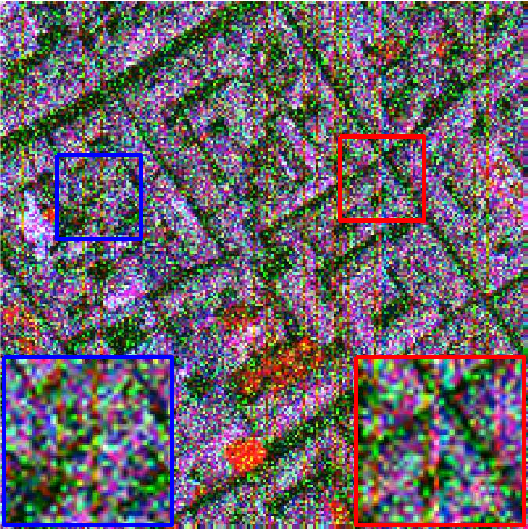}
  \caption{Noisy}
  \label{fig:pavia_noisy}
\end{subfigure}
\begin{subfigure}[b]{0.124\linewidth}
  \includegraphics[width=\linewidth]{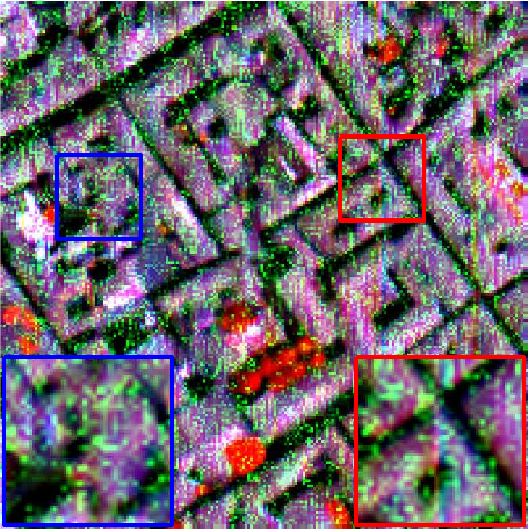}
  \caption{BM4D}
  \label{fig:pavia_BM4D}
\end{subfigure}
\begin{subfigure}[b]{0.124\linewidth}
  \includegraphics[width=\linewidth]{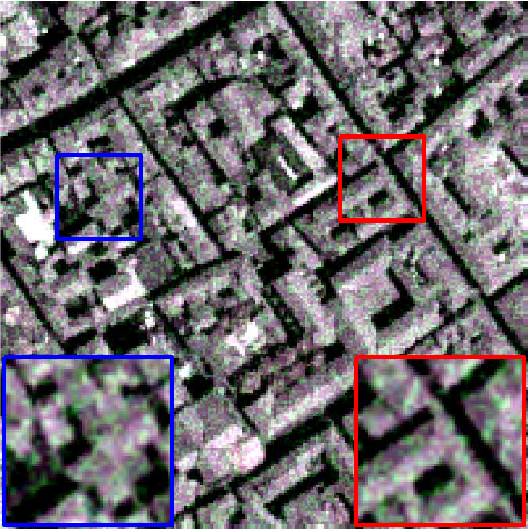}
  \caption{$LRTFL_0$}
  \label{fig:pavia_lrtfl0}
\end{subfigure}
\begin{subfigure}[b]{0.124\linewidth}
  \includegraphics[width=\linewidth]{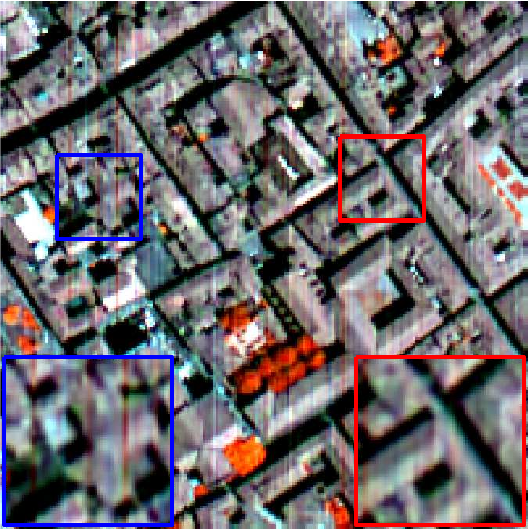}
  \caption{FastHyDe}
  \label{fig:pavia_FastHYDe}
\end{subfigure}
\begin{subfigure}[b]{0.124\linewidth}
  \includegraphics[width=\linewidth]{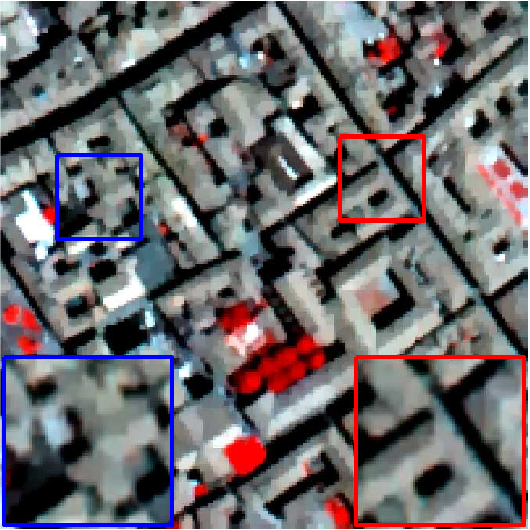}
  \caption{e3dtv}
  \label{fig:pavia_e3dtv}
\end{subfigure}
\begin{subfigure}[b]{0.124\linewidth}
  \includegraphics[width=\linewidth]{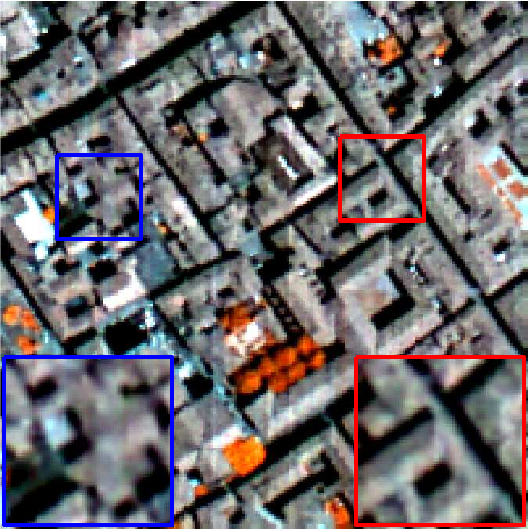}
  \caption{MAC-Net}
  \label{fig:pavia_MAC-Net}
\end{subfigure}
\begin{subfigure}[b]{0.124\linewidth}
  \includegraphics[width=\linewidth]{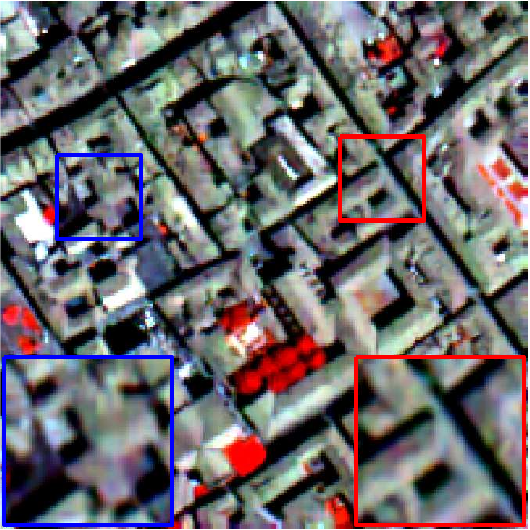}
  \caption{T3SC}
  \label{fig:pavia_t3sc}
\end{subfigure}
\begin{subfigure}[t]{0.124\linewidth}
  \includegraphics[width=\linewidth]{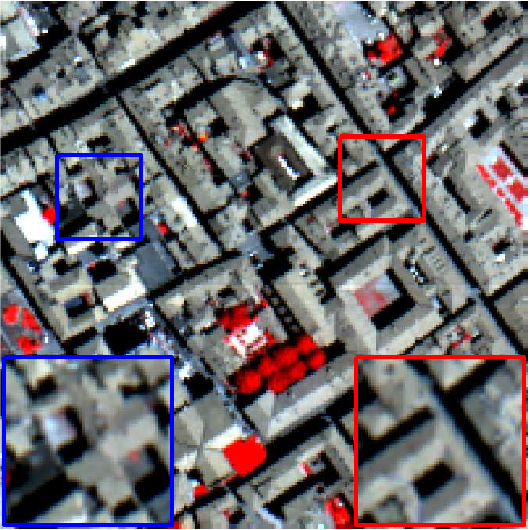}
  \caption{Clean}
  \label{fig:pavia_clean}
\end{subfigure}
\begin{subfigure}[t]{0.124\linewidth}
  \includegraphics[width=\linewidth]{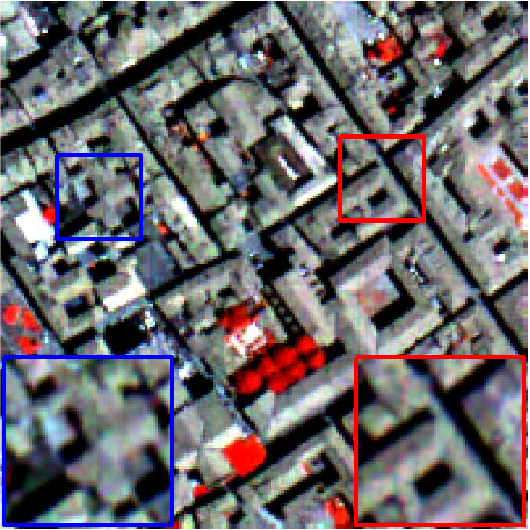}
  \caption{TRQ3D}
  \label{fig:pavia_TRQ3D}
\end{subfigure}
\begin{subfigure}[t]{0.124\linewidth}
  \includegraphics[width=\linewidth]{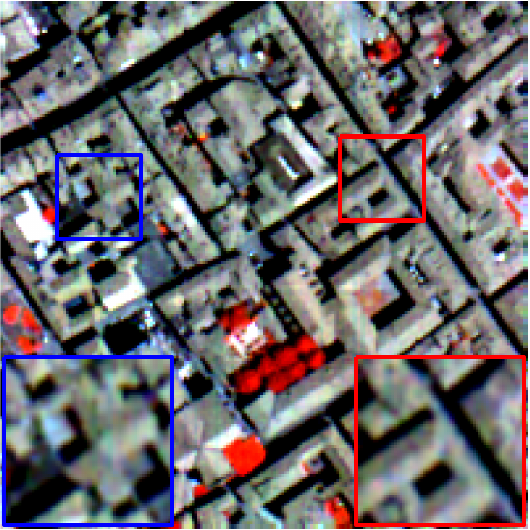}
  \caption{SERT}
  \label{fig:paiva_SERT}
\end{subfigure}
\begin{subfigure}[t]{0.124\linewidth}
  \includegraphics[width=\linewidth]{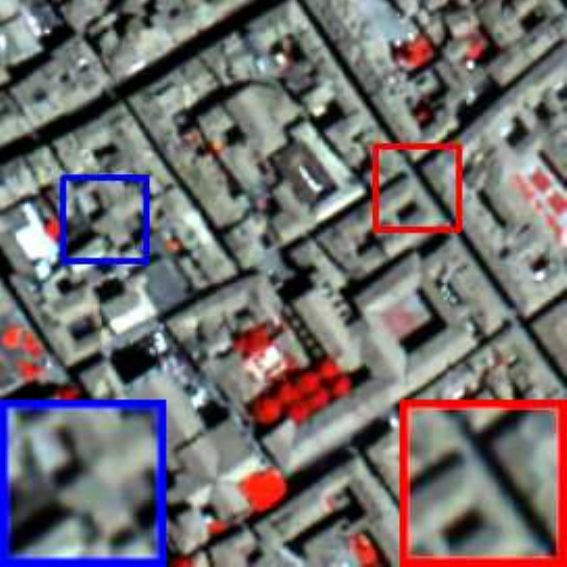}
  \caption{SSRT}
  \label{fig:pavia_SSRT}
\end{subfigure}
\begin{subfigure}[t]{0.124\linewidth}
\includegraphics[width=\linewidth]{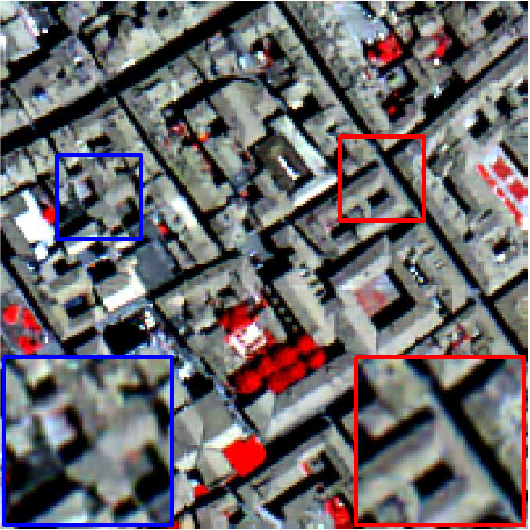}
  \caption{SSUMamba}
  \label{fig:pavia_ssumamba}
\end{subfigure}
\begin{subfigure}[t]{0.124\linewidth}
\includegraphics[width=\linewidth]{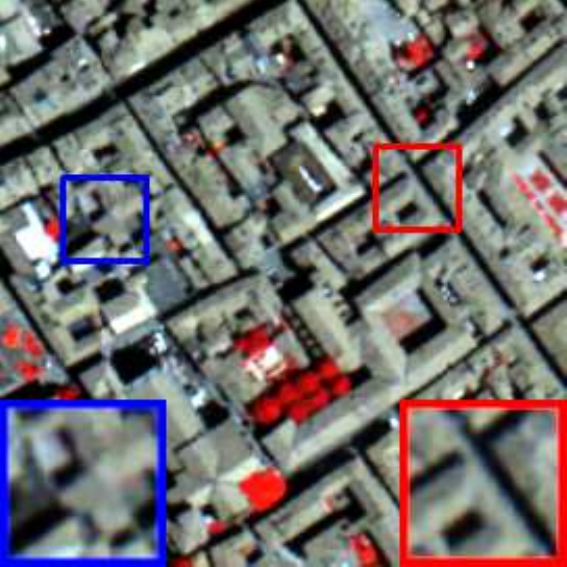}
  \caption{Proposed\\(5-mixture)}
  \label{fig:pavia_ensemble}
\end{subfigure}
\begin{subfigure}[t]{0.124\linewidth}
  \includegraphics[width=\linewidth]{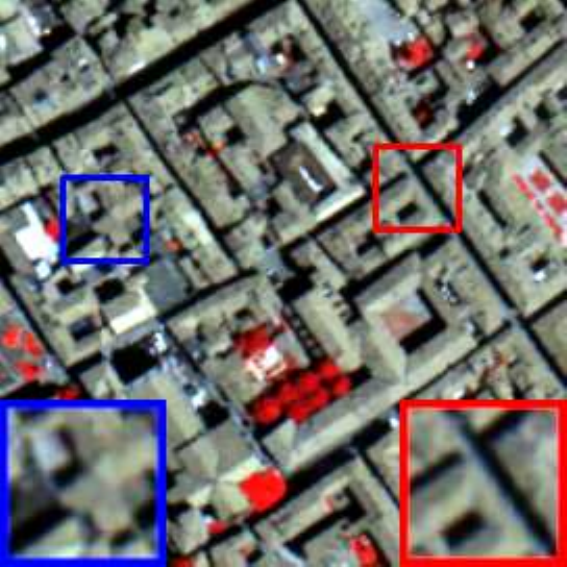}
  \caption{Proposed\\(1-single)}
  \label{fig:paiva-single}
\end{subfigure}
\caption{Denoising results on the Pavia city center with mixture noise, bands 65,45,25.}
\label{fig:pavia_visual}
\end{figure*}

\begin{figure*}[t]
  \captionsetup{font=small}
\captionsetup[sub]{font=scriptsize,skip=2pt}
\centering

\begin{subfigure}[t]{0.16\linewidth}
  \includegraphics[width=\linewidth]{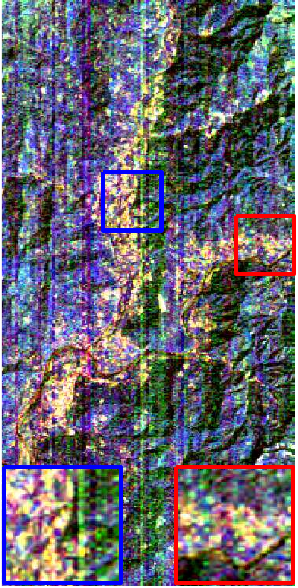}
  \caption{Noisy}
  \label{fig:eo1_noisy}
\end{subfigure}
\begin{subfigure}[t]{0.16\linewidth}
  \includegraphics[width=\linewidth]{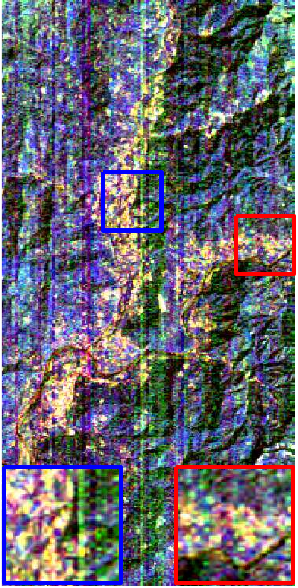}
  \caption{BM4D}
  \label{fig:eo1_BM4D}
\end{subfigure}
\begin{subfigure}[t]{0.16\linewidth}
  \includegraphics[width=\linewidth]{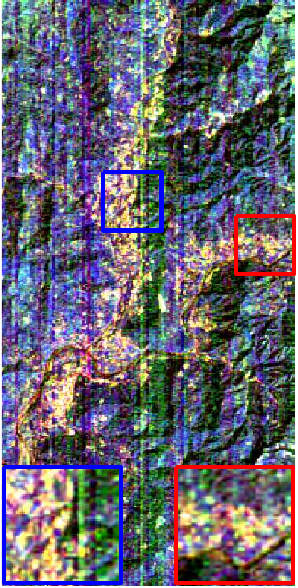}
  \caption{$LRTFL_0$}
  \label{fig:eo1_lrtfl0}
\end{subfigure}
\begin{subfigure}[t]{0.16\linewidth}
  \includegraphics[width=\linewidth]{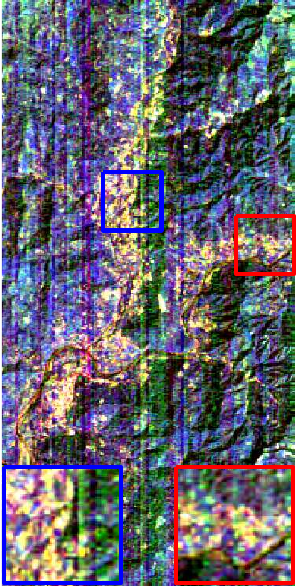}
  \caption{FastHyDe}
  \label{fig:eo1_FastHYDe}
\end{subfigure}
\begin{subfigure}[t]{0.16\linewidth}
  \includegraphics[width=\linewidth]{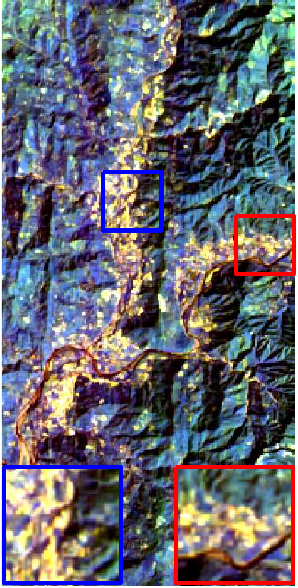}
  \caption{e3dtv}
  \label{fig:eo1_e3dtv}
\end{subfigure}
\begin{subfigure}[t]{0.16\linewidth}
  \includegraphics[width=\linewidth]{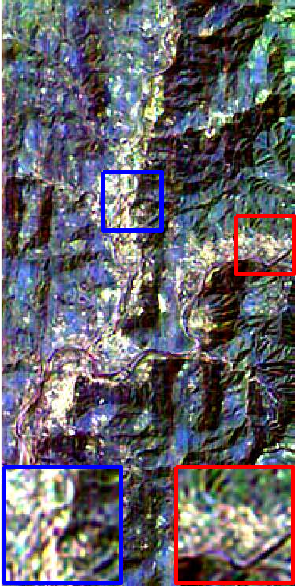}
  \caption{T3SC}
  \label{fig:T3SC}
\end{subfigure}

\par\vspace{2pt}

\begin{subfigure}[t]{0.16\linewidth}
  \includegraphics[width=\linewidth]{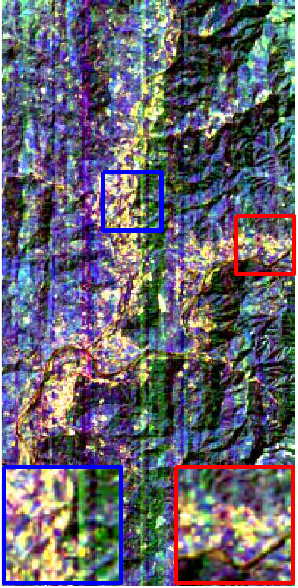}
  \caption{MAC-Net}
  \label{fig:eo1_MAC-Net}
\end{subfigure} 
\begin{subfigure}[t]{0.16\linewidth}
  \includegraphics[width=\linewidth]{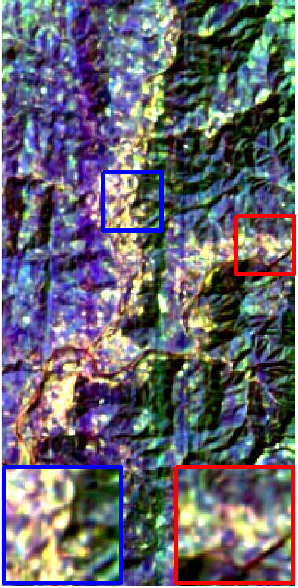}
  \caption{TRQ3D}
  \label{fig:eo1_TRQ3D}
\end{subfigure}
\begin{subfigure}[t]{0.16\linewidth}
  \includegraphics[width=\linewidth]{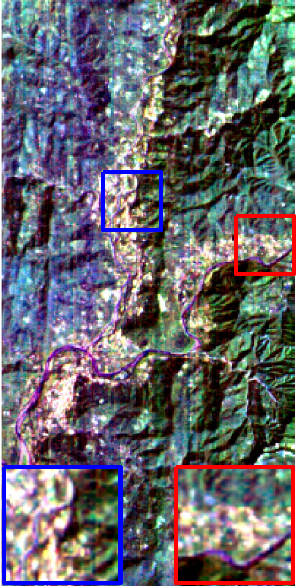}
  \caption{SERT}
  \label{fig:eo1_SERT}
\end{subfigure}
\begin{subfigure}[t]{0.16\linewidth}
  \includegraphics[width=\linewidth]{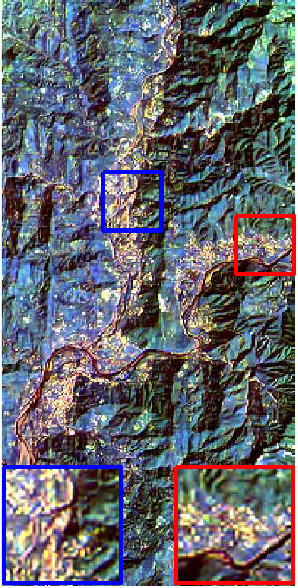}
  \caption{SSUMamba}
  \label{fig:eoi_ssumamba}
\end{subfigure}
\begin{subfigure}[t]{0.16\linewidth}
  \includegraphics[width=\linewidth]{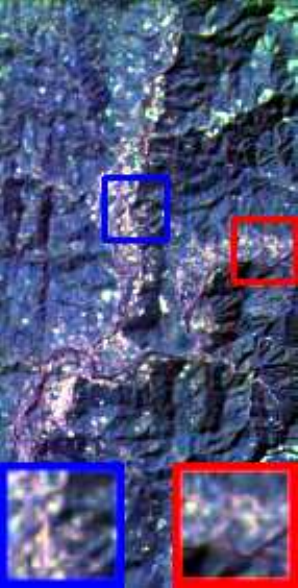}
  \caption{Proposed(5-mixture)}
  \label{fig:eo1_ensemble}
\end{subfigure}
\begin{subfigure}[t]{0.16\linewidth}
  \includegraphics[width=\linewidth]{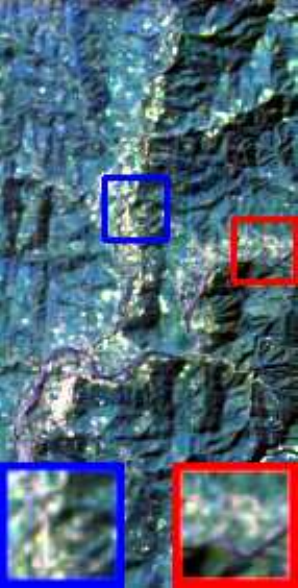}
  \caption{Proposed(1-single)}
  \label{fig:eo1_single}
\end{subfigure}

\caption{Denoising results on Earth Observing-1 HSI. The false-color images are generated by combining bands 163, 96, and 30}
\label{fig:eo1_visual}
\end{figure*}

\begin{figure*}[t]
  \captionsetup{font=small}
\captionsetup[sub]{font=scriptsize,skip=2pt}
\centering
\begin{subfigure}[t]{0.155\linewidth}
  \includegraphics[width=\linewidth]{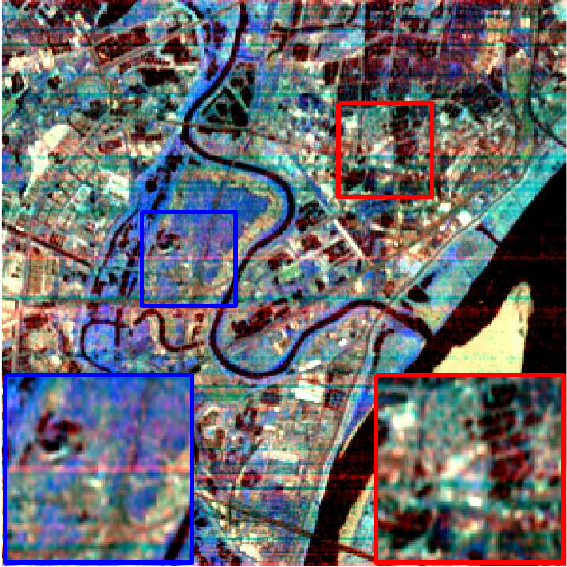}
  \caption{Noisy}
  \label{fig:wuhan_noisy}
\end{subfigure}
\begin{subfigure}[t]{0.155\linewidth}
  \includegraphics[width=\linewidth]{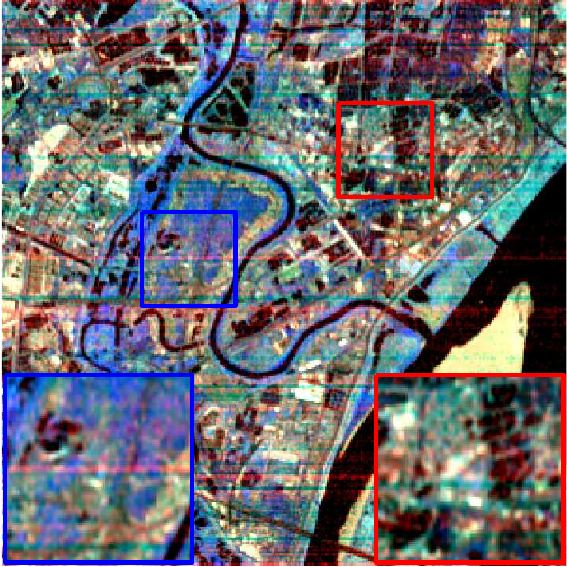}
  \caption{BM4D}
  \label{fig:wuhan_BM4D}
\end{subfigure}
\begin{subfigure}[t]{0.155\linewidth}
  \includegraphics[width=\linewidth]{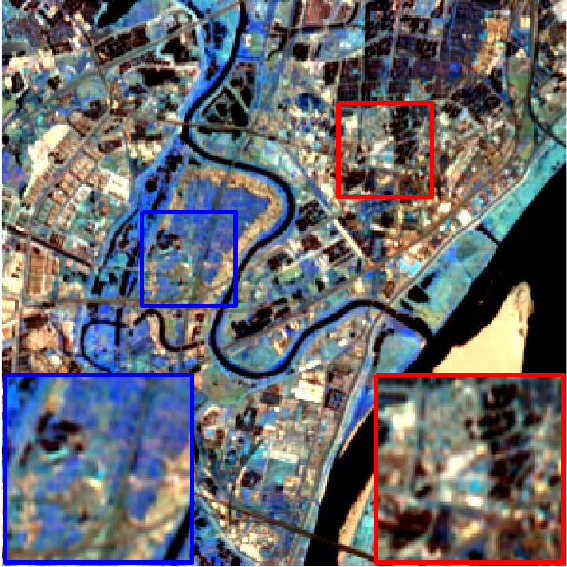}
  \caption{$LRTFL_0$}
  \label{fig:wuhan_lrtfl0}
\end{subfigure}
\begin{subfigure}[t]{0.155\linewidth}
  \includegraphics[width=\linewidth]{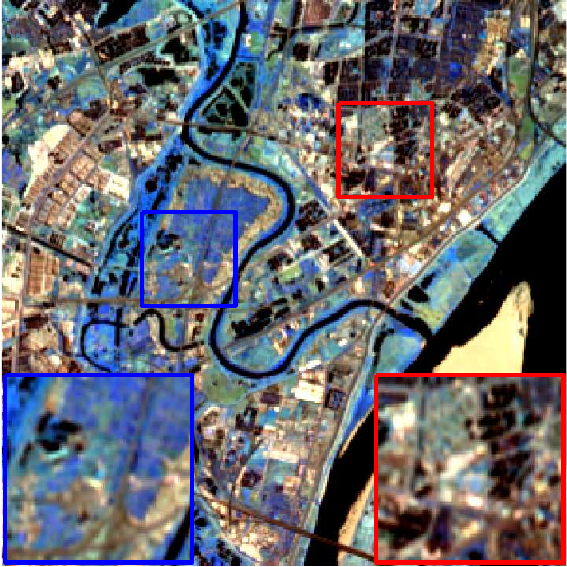}
  \caption{FastHyDe}
  \label{fig:wuhan_FastHYDe}
\end{subfigure}
\begin{subfigure}[t]{0.155\linewidth}
  \includegraphics[width=\linewidth]{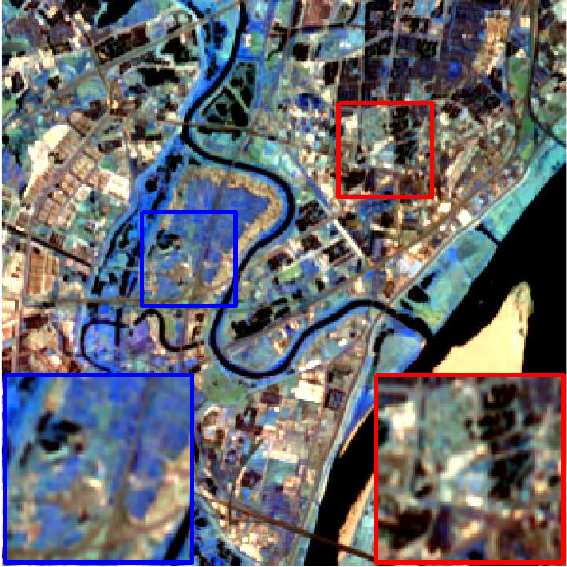}
  \caption{e3dtv}
  \label{fig:wuhan_e3dtv}
\end{subfigure}
\begin{subfigure}[t]{0.155\linewidth}
  \includegraphics[width=\linewidth]{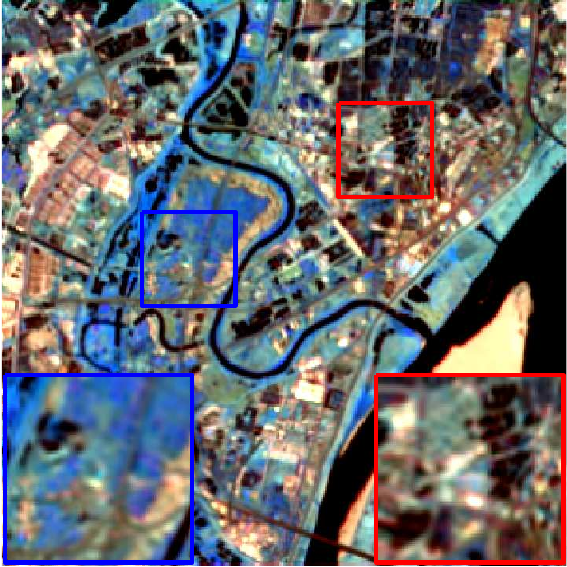}
  \caption{MAC-Net}
  \label{fig:wuhan_MAC-Net}
\end{subfigure}

\begin{subfigure}[t]{0.155\linewidth}
  \includegraphics[width=\linewidth]{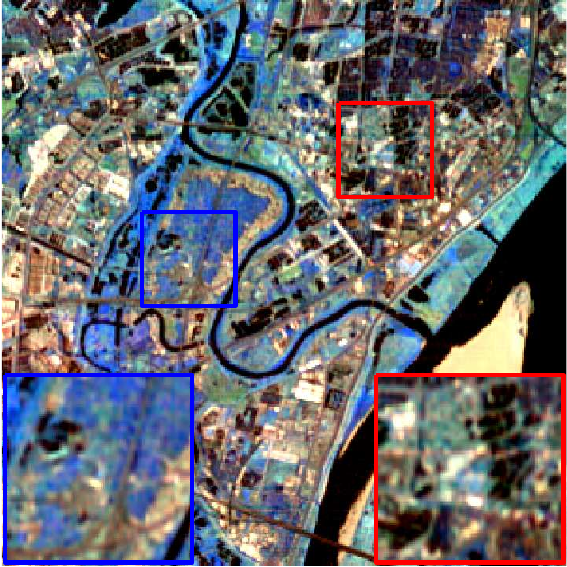}
  \caption{T3SC}
  \label{fig:wuhan_t3sc}
\end{subfigure}
\begin{subfigure}[t]{0.155\linewidth}
  \includegraphics[width=\linewidth]{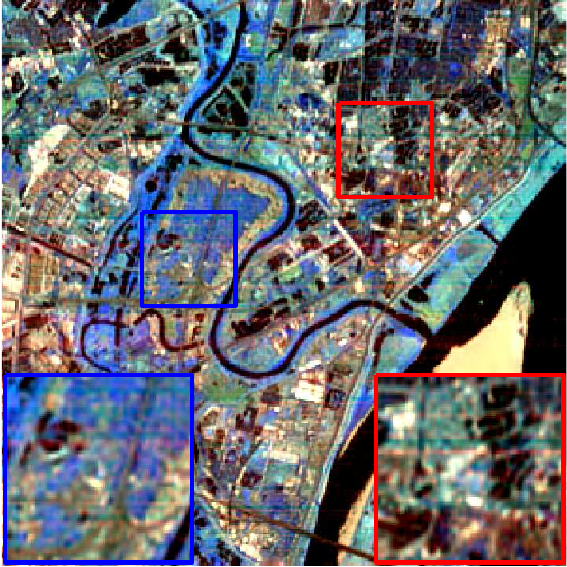}
  \caption{TRQ3D}
  \label{fig:wuhan_TRQ3D}
\end{subfigure}
\begin{subfigure}[t]{0.155\linewidth}
  \includegraphics[width=\linewidth]{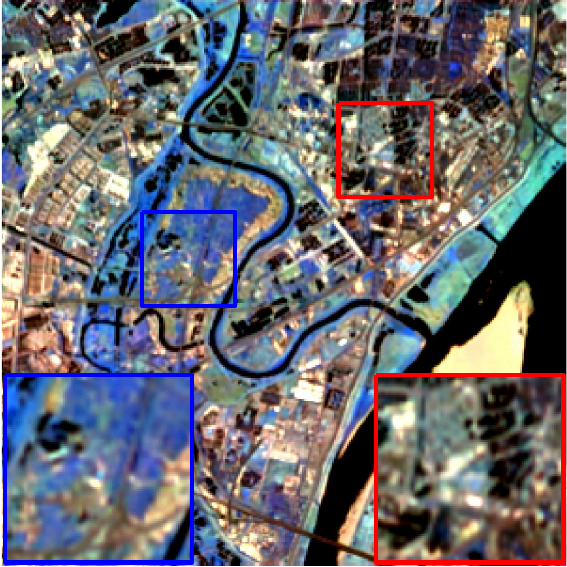}
  \caption{SERT}
  \label{fig:wuhan_SERT}
\end{subfigure}
\begin{subfigure}[t]{0.155\linewidth}
  \includegraphics[width=\linewidth]{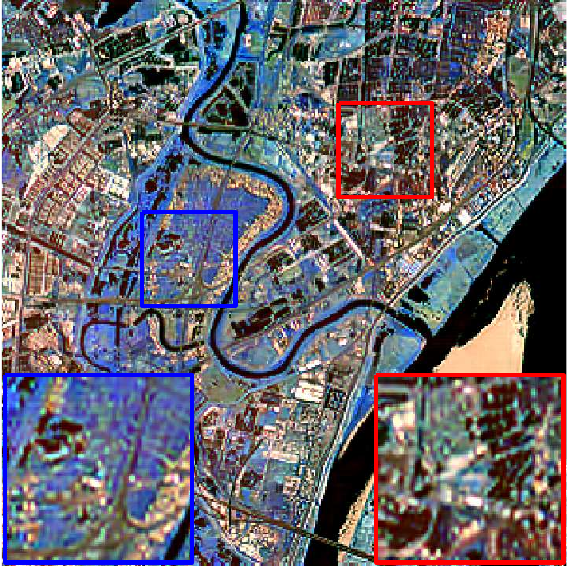}
  \caption{SSUMamba}
  \label{fig:wuhan_ssumamba}
\end{subfigure}
\begin{subfigure}[t]{0.155\linewidth}
  \includegraphics[width=\linewidth]{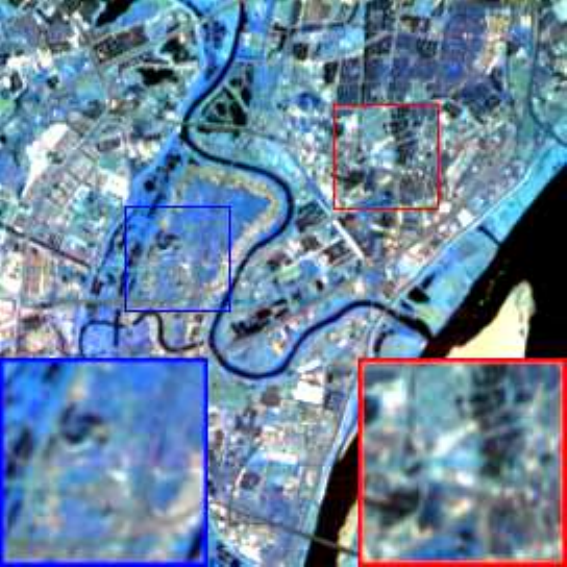}
  \caption{Proposed(5-mixture)}
  \label{fig:wuhan_ensemble}
\end{subfigure}
\begin{subfigure}[t]{0.155\linewidth}
  \includegraphics[width=\linewidth]{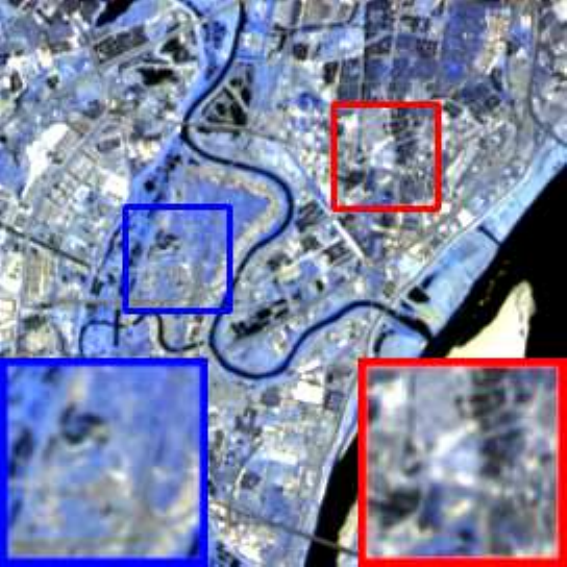}
  \caption{Proposed(1-single)}
  \label{fig:wuhan_single}
\end{subfigure}

\caption{Denoising results on the GF5-Wuhan HSI with mixture noise,The false-color images are generated by combining bands 152, 83, and 33.}
\label{fig:GF5_visual}
\end{figure*}


\begin{table*}[t]
\centering
\caption{Quantitative results on ICVL dataset under four noise conditions: Gaussian noise with $\sigma \in [0,15]$, $\sigma \in [0,55]$, $\sigma \in [0,95]$, and Mixture Noise. 
}
\label{tab:comparison}
\resizebox{\textwidth}{!}{
\begin{tabular}{lcccccccccccccc}
\toprule
\multirow{2}{*}{Method} & \multirow{2}{*}{Params (M)} & \multirow{2}{*}{FLOPs/pixel} 
& \multicolumn{3}{c|}{$\sigma \in [0,15]$} 
& \multicolumn{3}{c|}{$\sigma \in [0,55]$} 
& \multicolumn{3}{c|}{$\sigma \in [0,95]$} 
& \multicolumn{3}{c}{Mixture Noise} \\
\cmidrule(lr){4-6} \cmidrule(lr){7-9} \cmidrule(lr){10-12} \cmidrule(l){13-15}
 & & & PSNR↑ & SSIM↑ & SAM↓ & PSNR↑ & SSIM↑ & SAM↓ & PSNR↑ & SSIM↑ & SAM↓ & PSNR↑ & SSIM↑ & SAM↓ \\
\midrule
BM4D~\cite{6253256}         & -    & -      & 44.39 & 0.9683 & 0.0692 & 37.63 & 0.9008 & 0.1397 & 34.71 & 0.8402 & 0.1906 & 23.36 & 0.4275 & 0.5476 \\
LRTFL$_{0}$\cite{xiong2019hyperspectral}         & -    & -      & 43.41 & 0.9315 & 0.0570 & 35.63 & 0.8125 & 0.1914 & 32.83 & 0.7482 & 0.3014 & 30.93 & 0.8378 & 0.3613 \\
FastHyDe~\cite{zhuang2018fast}      & -    & -      & 48.08 & 0.9917  & 0.0404 & 42.86 & 0.9800 & 0.0630 & 40.84 & 0.9734 & 0.0771 & 27.58 & 0.7250 & 0.4534 \\
E-3DTV~\cite{peng2018enhanced}      & -    & -      & 46.05 & 0.9811& 0.0560 & 40.20 & 0.9505 & 0.0993 & 37.80 & 0.9297 & 0.1317 & 34.90 & 0.9041 & 0.1468 \\
FastHyMix\cite{zhuang2021fasthymix}
& -    & -      & 49.29 & 0.9929& 0.0350 & 43.25 & 0.9763 &  0.0764 & 40.69 & 0.9590 & 0.1078 &  27.08 & 0.6803 & 0.4688\\
\cmidrule[0.5pt](lr){1-15}
SimpleNet\cite{li2025aim}& 4.3M&13.5K&46.01 &0.9884&0.0400&41.91& 0.9751&0.0516&40.43&0.9682&0.0581&37.18&0.9381&0.0839\\

T3SC~\cite{bodrito2021trainable}         & 0.83M  & 0.93M    & 49.68 & 0.9912 & 0.0486 & 45.15 & 0.9810 & 0.0652 & 43.10 & 0.9734& 0.0747 & 34.09 & 0.9052 & 0.2340 \\
MAC-Net~\cite{9631264}      & 0.43M  & 14.8K    & 48.21 & 0.9915 & 0.0387 & 43.74 & 0.9768 & 0.0528 & 41.24 & 0.9577 & 0.0841 & 28.44 & 0.7393 & 0.4154 \\
TRQ3D~\cite{pang2022trq3dnet}     & 0.68M  & 0.44M    & 46.43 & 0.9878 & 0.0437 & 44.64  & 0.9840 & 0.0487 & 43.54 & 0.9806 & 0.0523 & 39.73 & 0.9491 & 0.0869 \\
SERT~\cite{li2023spectral}         & 1.9M  & 0.41M    & 50.18 & 0.9977 & 0.0278 & 46.34 & 0.9951 & 0.0373 & 44.47 & 0.9929 & 0.0447 & 39.13 & 0.9679 & 0.0963 \\
SSRT~\cite{fu2024hyperspectral}         & 10.2M  & 2.78M    & 52.12 & 0.9950 & 0.0225 & 47.85 & 0.9894 & 0.0319 & 46.20 & 0.9862 & 0.0375 & 42.52 & 0.9570 & 0.0986 \\
SSUMamba~\cite{fu2024ssumamba}      & 10.4M  & 23.6M    & 51.34 & 0.9946 & 0.0256 & 46.85 & 0.9882 & 0.0375 & 45.36 & 0.9853 & 0.0439 & 43.07 & 0.9726 & 0.0710\\
LaMamba~\cite{11177616} & 24.2M&11.5M &51.87&0.9949&0.0233&47.25&0.9888&0.0366&45.69&0.9856&0.0418&43.00&0.9726&0.0751 \\
\cmidrule[0.5pt](lr){1-15}
Proposed (5-mixture)  & 5.3M  & 100K    & 49.07 & 0.9915 & 0.0338 & 43.65 &  0.9803 & 0.0480 & 42.27 &  0.9751 & 0.0560 & 39.42 & 0.9468 & 0.0811\\
Proposed (1-large)  & 5.3M  & 100K   & 48.93 & 0.9917 & 0.0339 & 44.07 & 0.9816 & 0.0472 & 42.20 & 0.9741 & 0.0561 & 39.67 & 0.9506 & 0.0772\\
\bottomrule
\end{tabular}}
\vspace{-10pt}
\end{table*}

\subsection{Experimental Setting} 
For training, we use the ICVL dataset \cite{arad_and_ben_shahar_2016_ECCV}, which contains 204 hyperspectral images with a spatial resolution of $1392 \times 1300$ and 31 spectral bands. Following the protocol in \cite{wei20203}, we select 100 images for training. For evaluation, we employ both synthetic and real-world test sets to assess generalization and robustness. The synthetic data includes of 50 images from ICVL, Houston 2018 \cite{9812472}, and Pavia City Center \cite{audebert2019deep}. The real noise dataset includes images from Gaofen-5 Wuhan HSI \cite{liu2019advanced} and Earth Observing-1 HSI \cite{middleton2013earth}. To ensure fair comparison across different methods, we strictly adhere to the standardized noise generation procedure defined in \cite{fu2024hyperspectral}. In particular two main kinds of noise are considered: i) non-i.i.d. Gaussian Noise with three standard deviation ranges ($\sigma \in [0,15]$, $\sigma \in [0,55]$, and $\sigma \in [0,95]$); ii) mixture noise containing non-i.i.d. Gaussian noise with $\sigma \in [0,95]$, impulse noise applied to one-third of the bands with intensities ranging from $10\%$ to $70\%$, stripe noise affecting 5\% to 15\% of columns in one-third of the bands, deadline artifacts on 5\% to 15\% of columns in one-third of the bands. All noise types are applied to the ICVL dataset, while only mixture noise is used for the Houston 2018 and Pavia City Center datasets.
 
Before training the ensemble, we independently pretrain five of the proposed denoisers for 300 epochs using different random seeds for 400 epochs using $64 \times 64$ patches. We use $F=96$ for the each mixture denoiser and $F=220$ for the one large denoiser. Concerning Mamba, we set $K=4$, $E=1$ and $D=16$ as the kernel of causal convolution, the expansion and state factors, respectively. We use the Adam optimizer with $\beta_1 = 0.9$, $\beta_2 = 0.999$, and an initial learning rate of $5 \times 10^{-4}$, which is halved at epoch 30 and then every subsequent 100 epochs. For synthetic datasets, performance is quantitatively assessed using:
Peak Signal-to-Noise Ratio (PSNR), Structural Similarity Index Measure (SSIM), Spectral Angle Mapper (SAM).
For real-noise datasets, we adopt two no-reference image quality metrics, TOPIQ NR \cite{chen2024topiq}, and CLIPIQA+ \cite{wang2022exploring}. Training has been performed on a single A40 GPU and requires approximately 5 days. Code will be released at \url{https://github.com/diegovalsesia/scalable-pushbroom-architectures}.

\begin{table*}[t]
\centering
\caption{Quantitative comparison of HSI denoising methods on other datasets}
\label{tab:otherdataset}
\resizebox{\textwidth}{!}{
\begin{tabular}{lccccccccccccc}
\toprule
\multirow{2}{*}{Method} & \multirow{2}{*}{Params (M)} & \multirow{2}{*}{FLOPs/pixel} 
& \multicolumn{3}{c|}{Huston 2018} 
& \multicolumn{3}{c|}{PAVIA CITY CENTER}  & \multicolumn{2}{c|}{GAOFEN-5 WUHAN} & \multicolumn{2}{c}{EARTH OBSERVING-1}\\
\cmidrule(lr){4-6} \cmidrule(lr){7-9}\cmidrule(lr){10-11}\cmidrule(lr){12-13}
 & & & PSNR↑ & SSIM↑ & SAM↓ & PSNR↑ & SSIM↑ & SAM↓ &TOPIQ NR↑&CLIPIQA+↑ &TOPIQ NR↑&CLIPIQA+↑ \\
\midrule
BM4D~\cite{6253256}         & -    & -      & 22.76 & 0.4762 & 0.5168 & 21.70 & 0.5128 & 0.5297 &0.3828 &0.5827 &0.5184&0.5011\\
LRTFL$_{0}$\cite{xiong2019hyperspectral}       & -    & -      & 28.75 & 0.8038 & 0.2221 & 26.49  & 0.8147 & 0.3703 &0.3877 &0.6021&0.5177&0.5204&\\
FastHyDe~\cite{zhuang2018fast}      & -    & -      & 27.07& 0.7757  & 0.4518 & 26.78 & 0.8361 & 0.4040 &0.3887&0.6021&0.5235&0.5187\\
E-3DTV~\cite{peng2018enhanced}      & -    & -      & 30.64 & 0.8570& 0.1323 & 30.44 & 0.8941 & 0.1134&0.3815&0.6055&0.4838&0.5355\\
FASTHyMix~\cite{zhuang2021fasthymix} &-&-& 25.35&0.7131&0.4593 &26.52&0.7982&0.4151 &0.3894 & 0.2704& -&-\\
\cmidrule[0.5pt](lr){1-13}

Simplenet~\cite{li2025aim} &4.3M & 13.5K & 30.03&0.8781 &0.1832 & 29.36  &0.8484 &0.2278 &0.3628 & 0.5444 &0.4552&0.4904  \\
T3SC~\cite{bodrito2021trainable}         &  0.83M  & 0.93M    & 29.84 & 0.8751 & 0.1943 & 28.69 & 0.8656 & 0.2135 &0.4274&0.5735&0.5258&0.4892\\
MAC-Net~\cite{9631264}      & 0.43M  & 14.8K    & 28.83 & 0.7963 & 0.2356 & 27.74 & 0.8724 & 0.3222 &0.3477 &0.5549&0.5231&0.4964&  \\
TRQ3D~\cite{pang2022trq3dnet}     & 0.68M  & 0.44M    & 32.55 & 0.9194 & 0.1241 & 28.23 & 0.8665 & 0.1961 &0.3840&0.5956 &0.3536&0.4437 \\
SERT~\cite{li2023spectral}         & 1.9M  & 0.41M    & 31.31 & 0.9296 & 0.1517 & 32.16 & 0.9544 & 0.1384 &0.3824&0.6081 &0.4654&0.5161 \\
SSRT~\cite{fu2024hyperspectral}         & 10.2M  & 2.78M    & 34.71 &0.9387 & 0.0915 & 35.10 & 0.9493 & 0.1036 &-&-&-&- \\
SSUMamba~\cite{fu2024ssumamba}      & 10.4M  & 23.6M    & 34.74 & 0.9452 & 0.0993 & 35.70 & 0.9546 & 0.1057 &0.4969 &0.6481 &0.6074 &0.5559\\
LAMamaba~\cite{11177616} & 24.2M &11.5M & 35.10&0.9462&0.0950 &36.67&0.9634&0.1000& -&-&0.5853&0.5092 \\
\cmidrule[0.5pt](lr){1-13}
Proposed (5-mixture)  & 5.3M  & 100K    & 30.39 & 0.8843 & 0.1577 & 33.23 & 0.9202 & 0.1808&0.3902&0.5940& 0.5114&0.5184\\
Proposed (1 large)  & 5.3M  & 100K    & 29.86 &  0.8753 & 0.1772 & 33.30 & 0.9173 & 0.1854 &0.3824 &0.5992 &0.5026 & 0.5161 \\
\bottomrule
\end{tabular}}
\end{table*}

\subsection{Quantitative comparisons with state of the art}

In this section, we compare our proposed method against a broad spectrum of state-of-the-art HSI denoising techniques, ranging from traditional model-based approaches, including BM4D\cite{6253256}, $\text{LRTFL}_{0}$ \cite{xiong2019hyperspectral}, FastHyDe \cite{zhuang2018fast}, and E-3DTV \cite{peng2018enhanced}, to recent deep learning methods. Concerning the latter, we evaluate convolutional models such as T3SC \cite{bodrito2021trainable}, 3D convolutional networks like MAC-Net \cite{9631264}, transformer-based architectures including TRQ3D \cite{pang2022trq3dnet}, SERT \cite{li2023spectral}, and SSRT \cite{fu2024hyperspectral}, as well as the SSM-based SSUMamba \cite{fu2024ssumamba} and LaMamba~\cite{11177616}. Moreover, a vary simple and lightweight baseline (SimpleNet) capable of real-time operations was adapted from SIDUNet, a mobile-optimized denoiser originally developed for RAW image enhancement in the AIM 2025 challenge \cite{li2025aim}.

Our experiment tests the best achievable quality by the proposed architecture, without power scalability tradeoffs or faults. In addition to the results of a mixture of 5 denoisers, we also benchmark the line-wise processing concept by reporting the performance of an individual denoiser with the same complexity as the mixture. Tables \ref{tab:comparison} and \ref{tab:otherdataset} show the complexity of the tested models as well as image quality under the various noise settings. We remark that the proposed approach is significantly less complex than existing methods at only 100K FLOPs/pixel, as we seek real-time runtime on low-power hardware, as later validated in Fig. \ref{fig:runtime}. Nevertheless, the proposed approach offers competitive image quality, sometimes even outperforming recent more complex designs. Fig. \ref{fig:houston_visual}, \ref{fig:pavia_visual}, \ref{fig:eo1_visual} and \ref{fig:GF5_visual} shows a qualitative comparison of denoised images. 

\subsection{Test on low-power hardware}
We now study the computational efficiency of the proposed method with respect to the state of the art. 

\begin{figure*}[t]
  \begin{minipage}[c]{0.33\textwidth}
    \vspace*{\fill}
    \centering
    \subcaptionbox{Lines\label{fig: line}}[\textwidth]{
      \includegraphics[width=\linewidth]{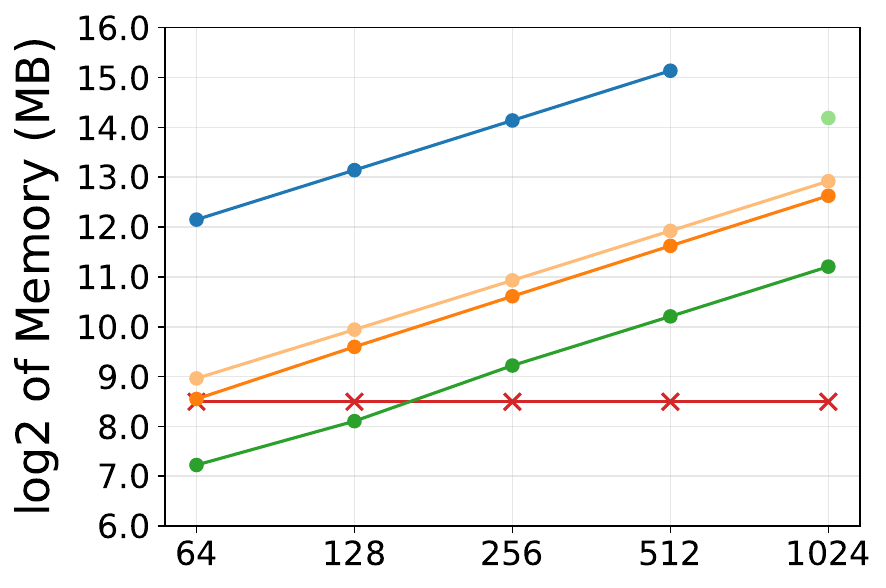}
    }
  \end{minipage}
  \hfill
  \begin{minipage}{0.33\textwidth}
    \subcaptionbox{ Columns \label{fig:Column}}[\textwidth]{
      \includegraphics[width=\linewidth]{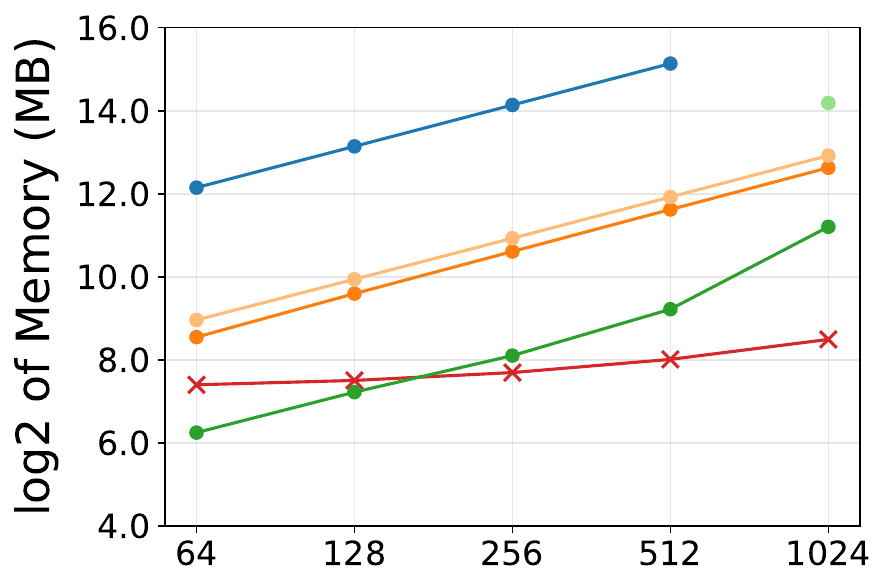}
    }
    \end{minipage}
    \hfill
    \begin{minipage}{0.33\textwidth}
    \subcaptionbox{ Bands \label{fig:bands}}[\textwidth]{
      \includegraphics[width=\linewidth]{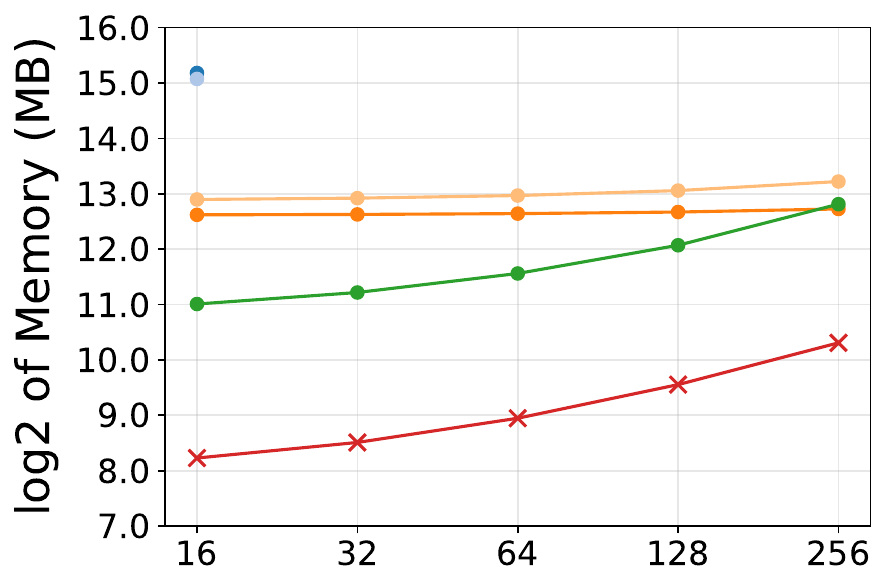}     
      }
    \end{minipage}
  \begin{minipage}{\textwidth}
    \centering
    \includegraphics[width=0.9\linewidth]{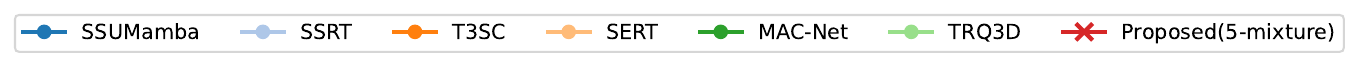}
  \end{minipage}
  \vspace{-10pt}
  \caption{Memory usage comparison across three dimensions: number of lines, and columns, spectral bands. Image size is a) $N_l \times 1024 \times 31$, b) $1024 \times N_c \times 31$, c) $1024 \times 1024 \times N_b$. Note that SSRT and TRQ3D support only square inputs and TRQ3D only supports a 31-band input. SSRT and SSUMamba run out of memory when the number of bands reaches 32.}
  \label{fig:memory}
\end{figure*}

Fig. \ref{fig:memory} highlights the superior memory efficiency of our method and its scaling as function of input lines, columns, and bands, showing that large images can be efficiently processed onboard. Notably, thanks to the line-wise design our method has constant memory requirements for any number of acquired lines.

Additionally, we tested runtime on a low-power system (Nvidia Jetson Orin Nano - 15W) that could be considered representative of onboard capabilities. All the architectures have been reduced to 16-bit floating point precision for both weights and activations. We use the Line Acquisition Time of the PRISMA satellite, which is 4.34 ms to acquire a line of size $1 \times 1000 \times 66$ as a reference for real time performance. Results are shown in Table \ref{tab:speed}. The table\footnote{We remark that the SSUMamba \cite{fu2024ssumamba} method is not present in the Table \ref{tab:speed} because making the authors' model run on the Jetson platform would require major code modifications.} shows the average time to process one line of size $1 \times 1000 \times 66$. For methods not working on a line-by-line basis we process an image of size $1000 \times 1000 \times 66$ and divide the total runtime by the number of lines. Fig. \ref{fig:runtime} presents the results as a quality against runtime tradeoff.
We can see that only the proposed method is capable of achieving the real-time processing requirement of 4.34 ms. These results confirm that, while theoretically providing slightly better image quality, state-of-the-art algorithms are unsuitable for onboard usage as they require impractically large amounts and memory and cannot run in real time, constituting a bottleneck, and the proposed approach is the only suitable for the task.

\subsection{Power Scalability}

\begin{figure*}[t]
  \centering
  \begin{minipage}[t]{0.48\textwidth}
    \centering
    \includegraphics[width=\textwidth]{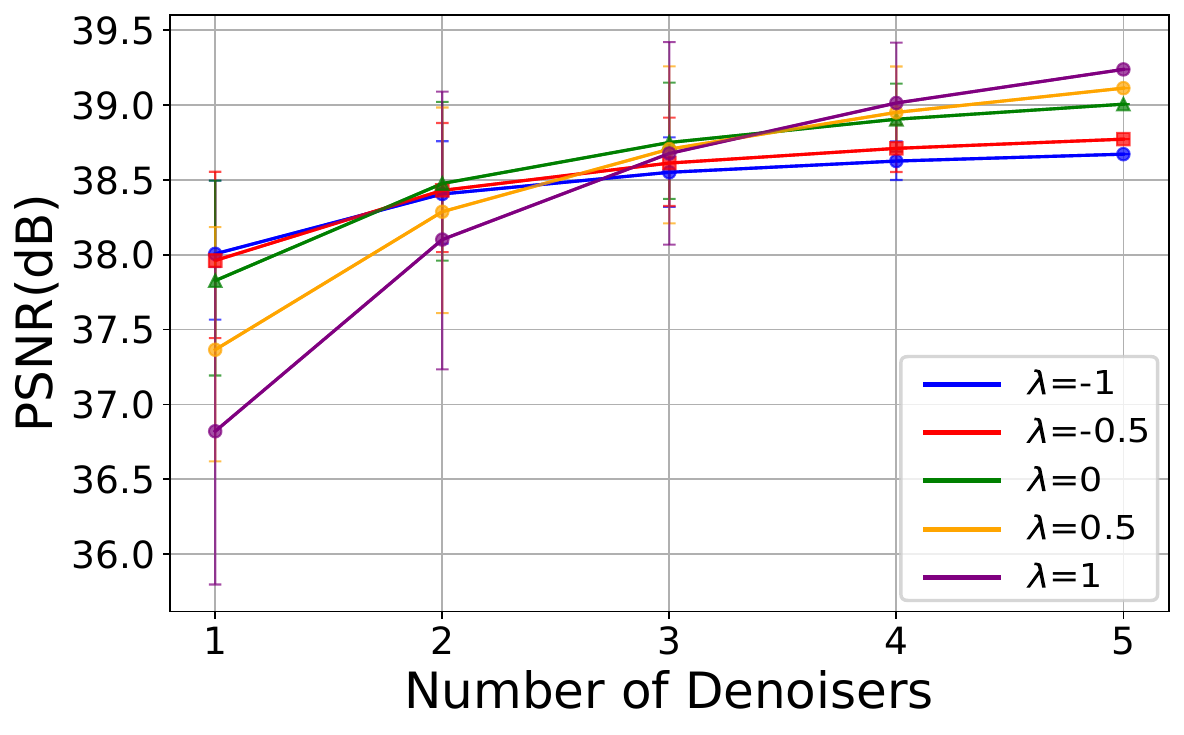}
    \vspace{-15pt}
    \caption{Power scalability tradeoffs. Power scalability factor $\lambda$ controls prioritization of high-power or low-power regimes.}
    \label{fig: power scalability}
  \end{minipage}
  \hfill
  \begin{minipage}[t]{0.48\textwidth}
    \centering
    \includegraphics[width=\textwidth]{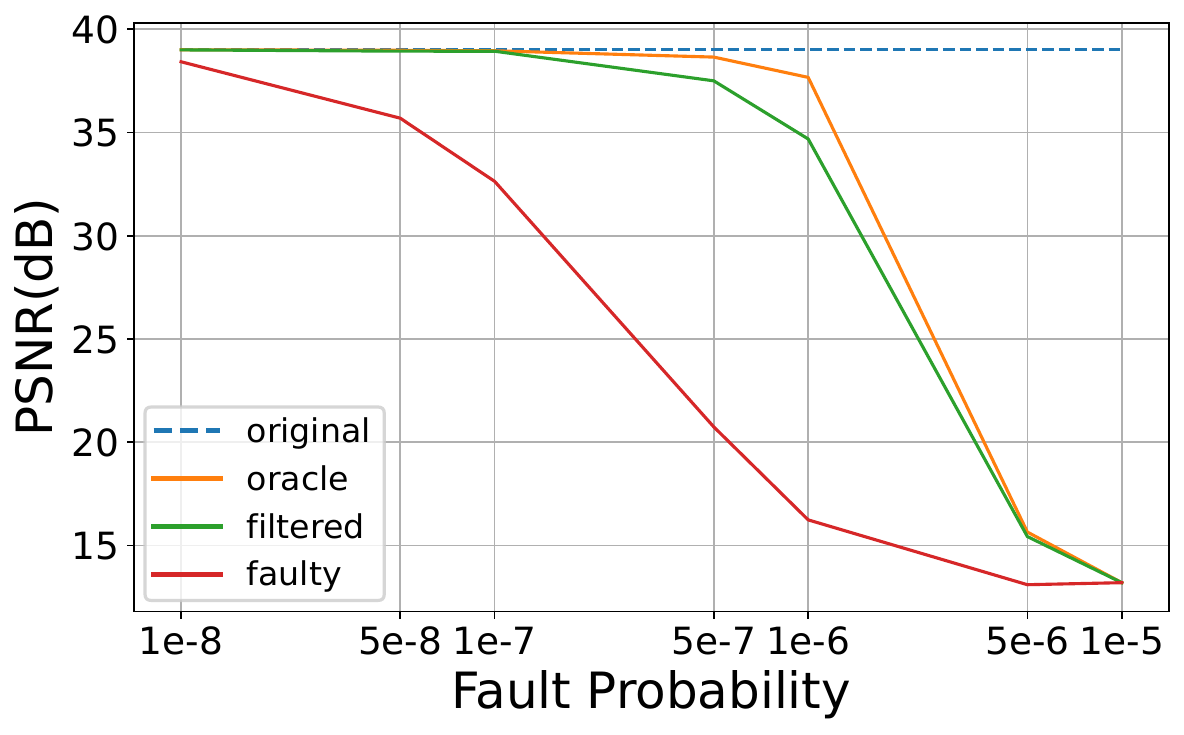}
    \vspace{-15pt}
    \caption{Output PSNR for the original model (no faults), the oracle fault detector, the proposed fault detector, and the unfiltered (faulty) mixture.}
    \label{fig:fault_injection}
  \end{minipage}
  \vspace{-15pt}
\end{figure*}

We now analyze the power scalability and fault resilience properties of the proposed design. Concerning power scalability, we trained our mixture model using different power scaling factors $\lambda\in\{-1,-0.5,0,0.5,1\}$ on the ICVL training set with a mixture noise pattern. The results are presented in Fig. \ref{fig: power scalability}, where the solid curves denotes the average performance and the error bars denote the variability due to the specific selection of subset of denoisers. We can observe that models trained with smaller $\lambda$ values exhibit better quality at low power (fewer denoisers used) compared to models where high $\lambda$ values are used, which conversely perform better in the high power regimes. This offers different tradeoffs to the system designer based on the expected distribution of power over time.
Specifically, these models perform better when only a single denoiser is used and show a steeper improvement as the number of denoisers increases. However, this improved scalability sacrificed peak performance, as models with lower $\lambda$ values tend to underperform when the full ensemble is utilized.

\subsection{Fault Resilience}
\label{sec:experiment_fault}
Regarding fault resilience, we use the PyTorchFI \cite{PytorchFIMahmoudAggarwalDSML20} library to simulate hardware faults by randomly injecting errors into the model weights. Focusing specifically on 1D convolutional and linear layers. Across a single denoiser, a total of 817,920 weights are subject to fault injection.
Fig. \ref{fig:fault_injection} shows how the PSNR of the denoised image degrades as a function of the probability of perturbing a weight, when no fault detection is performed, when faults are detected and filtered with the method in Sec. \ref{sec:fault} and with an oracle fault detector that exactly knows which denoisers are faulty. We can notice that the proposed method is quite robust and performs close to the oracle method, which demonstrates the output quality degradation as the fault probability increases. We use a threshold of $0.01$ because it gives the best trade-off between detecting faulty denoisers (high true positive rate) and avoiding false alarms (low false positive rate). Notably, the proposed fault detection and pruning strategy 
significantly mitigates the decline in denoising performance, highlighting the robustness of our ensemble model under fault-prone conditions.

\begin{figure*}
    \centering
    {\includegraphics[width=\linewidth]{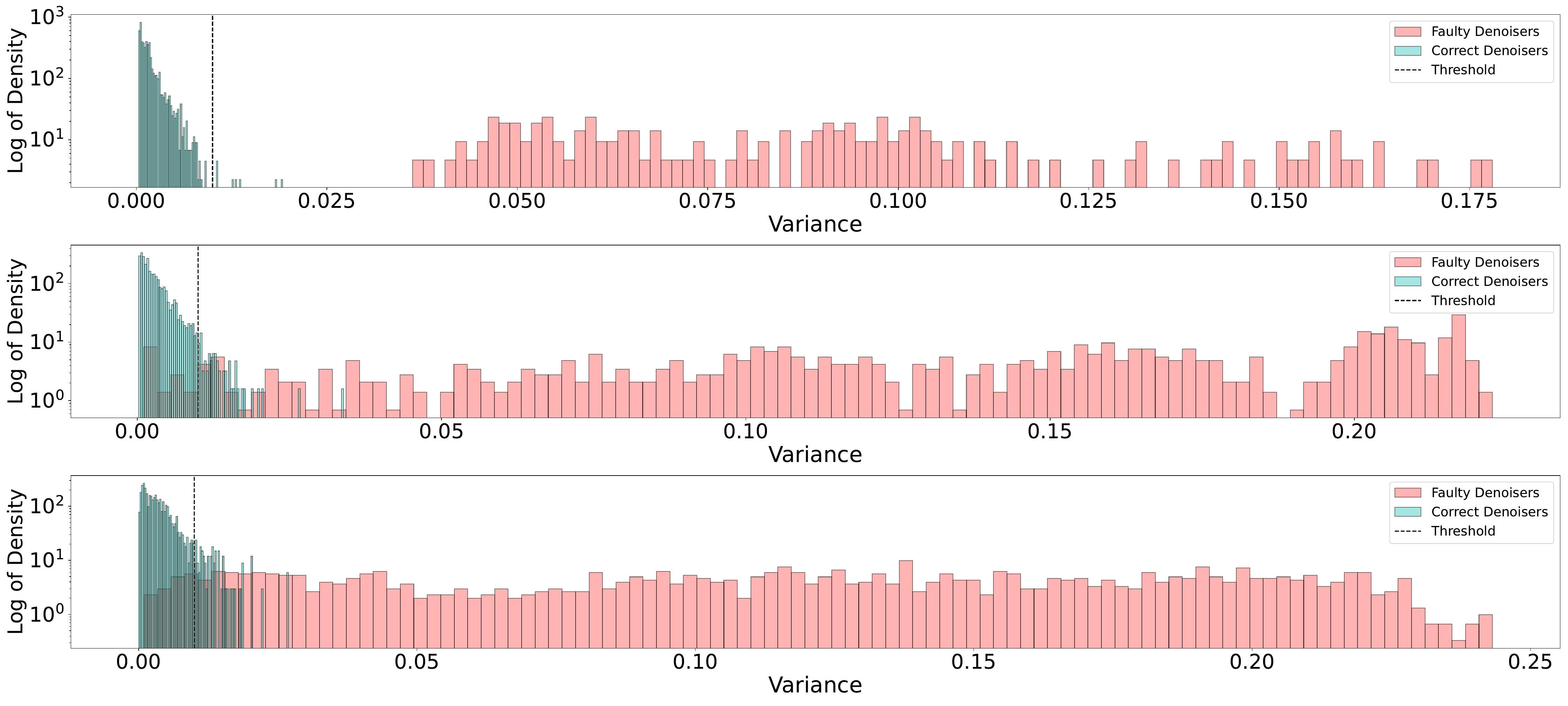}}
    \caption{Variance density distribution with probabilities $1\times10^{-7}$, $5\times10^{-7}$,and $1\times10^{-6}$ }
    \label{fig:variance}

  \vspace{-5pt}
\end{figure*}
In addition to showing that the proposed fault detection approach leads to superior image quality in presence of faults, we perform a statistical analysis to validate our hypothesis that the spatial variance of diagonal attention values does increase in presence of faults. To this end, we conducted an additional experiment in which faults were randomly injected into model weights with probabilities $1\times10^{-7}$, $5\times10^{-7}$, and $1\times10^{-6}$. For each probability level, we performed 30 independent trials and computed the spatial variances of diagonal attention values for both non-faulty and faulty denoisers on each test image in testset. The resulting distributions are shown in Fig.\ref{fig:variance}.

Across all fault probabilities, we observe that correct denoisers exhibit tightly clustered variance with lower values, indicating stable and coherent attention behavior. Instead, faulty denoisers produce significantly larger variances, forming a distinct distribution that is well separable from the one of non-faulty denoisers. The two distributions overlap more as the fault probability increasing, rendering detection more difficult, as expected.

Since this analysis has been conducted on a large number of images with several fault realizations, it is quite robust in showing evidence that variance the spatial variance of diagonal attention values does increase in presence of faults and its thresholding can be an effective method for fault filtering.

\subsection{Ablation of memory mechanism}
We tested the proposed method (1-large configuration) with other memory mechanisms, namely Causal Convolution and LSTM layers, in place of Mamba layers. Specifically, we fixed all the rest of the architecture and every hyperparameter and just substituted the two Mamba layers first with two Causal Convolution with kernel $4$ or with two LSTM layers having hidden the dimension equal to the model feature dimension. This allowed for the preservation of the line based acquisition system. We trained the Causal Convolution and LSTM models from scratch, employing the same training procedure used for the Mamba model. The results are shown in Table \ref{tab:complexity}. As expected, the selectivity of Mamba yields better performance whilst requiring fewer computational resources.  

\begin{table}[t]
\centering
\caption{Inference speed comparison.}
\label{tab:speed}
\begin{tabular}{ccc}
\textbf{Methods} & \textbf{Inference Time} & \textbf{PSNR} \\
\hline
\hline
{T3SC } & 9.76 ms & 34.09 dB\\
{MAC-Net} & 13.07 ms & 28.44 dB\\
{TRQ3D } & 28.28 ms & 39.73 dB\\
{SERT} & 20.19 ms & 39.13 dB\\
{SimpleNet} & 4.80 ms & 37.18 dB\\
{\textbf{Ours}}  &  \textbf{4.54 ms} & \textbf{39.42 dB} \\
\hline
\end{tabular}
\end{table}

\begin{table}[t]
\centering
\caption{Comparison with other sequence modeling architectures. Results obtained on ICVL with mixture of noise.}
\label{tab:complexity}
\begin{tabular}{ccc}
\textbf{Model} & \textbf{PSNR} & \textbf{FLOPs/pixel} \\
\hline
\hline
{CausalConv } & 39.17 dB & 121.6 K \\
{LSTM} & 39.52 dB & 145.6 K \\
{\textbf{Mamba}}  &  \textbf{39.67 dB} & \textbf{100 K} \\
\hline
\end{tabular}
\end{table} 

\subsection{Sensitivity to Number of Denoisers \textit{D}}
\begin{figure*}[t]
  \centering
  \begin{minipage}[t]{0.48\textwidth}
    \centering
    \includegraphics[width=\textwidth]{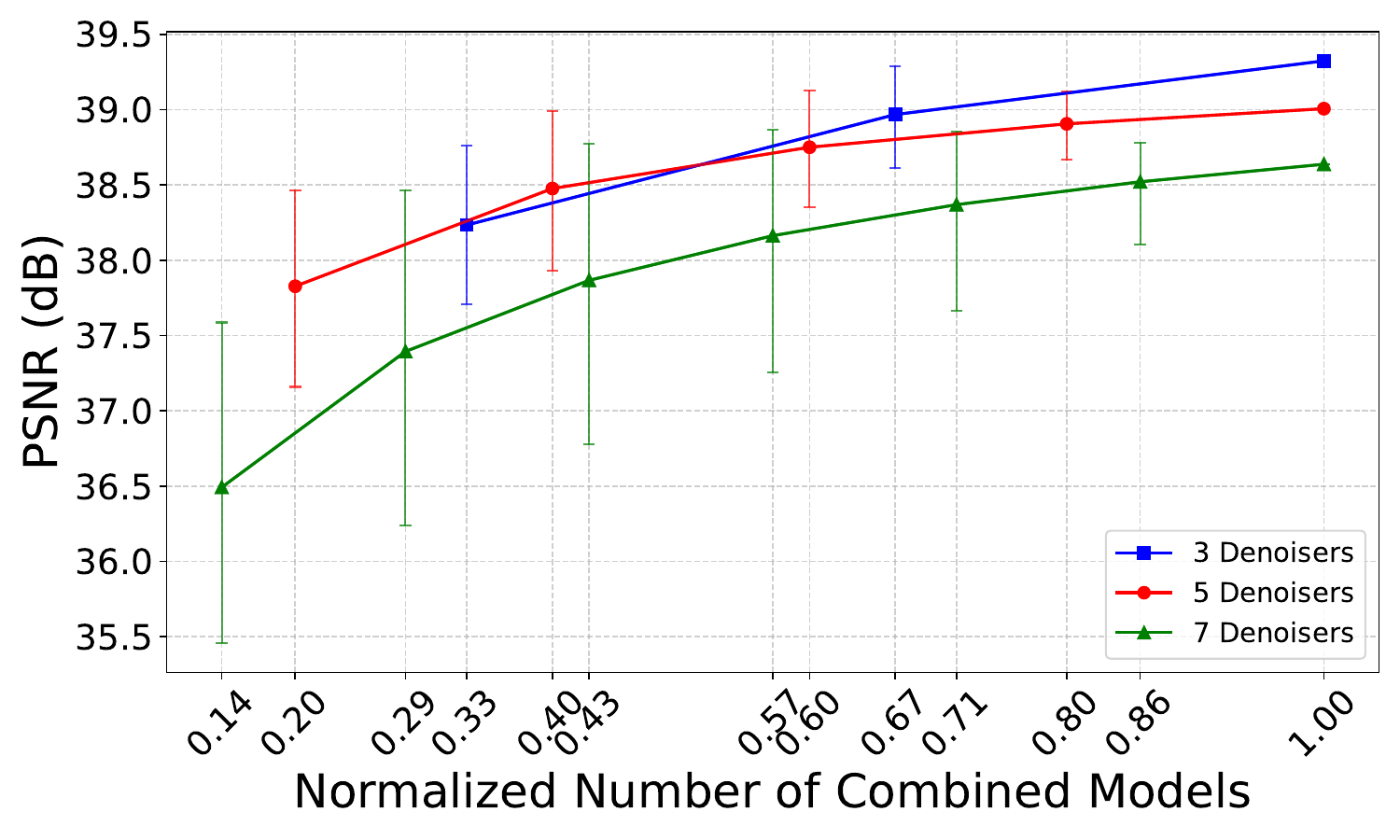}
    \vspace{-15pt}
    \caption{Sensitivity of Power Scalability to total number of denoisers $D$. The x axis can be interpreted as a percentage of maximum power.}
    \label{fig: diff_denoiers_power}
  \end{minipage}
  \hfill
  \begin{minipage}[t]{0.48\textwidth}
    \centering
    \includegraphics[width=\textwidth]{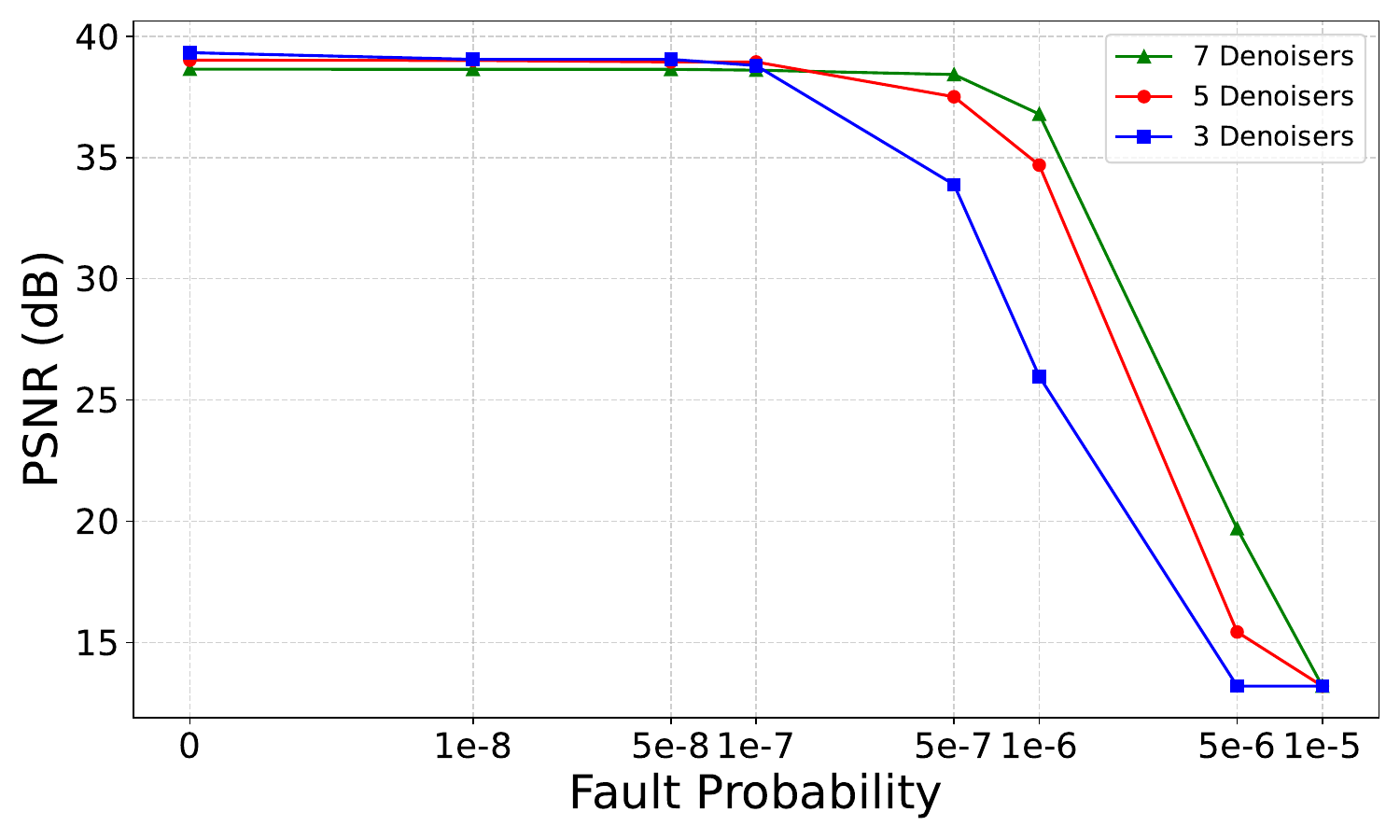}
    \vspace{-15pt}
    \caption{Sensitivity of fault detection to total number of denoisers $D$.}
    \label{fig:diff_denoisers_fault_injection}
  \end{minipage}
  \vspace{-15pt}
\end{figure*}

Different numbers of denoisers lead to different tradeoffs in terms of peak performance, power scalability, and fault resilience. For example, using too few denoisers reduces the granularity of power scaling and limits fault tolerance, while using too many can reduce peak performance and increase power requirements due to the need for additional accelerators. Our main results adopted five denoisers ($D=5$) as a balanced configuration that offers a good trade-off among performance, scalability, and robustness in our baseline experiments.
In this section, we investigate the sensitivity of such properties to different numbers of denoisers, by also considering a mixture with a small number of denoisers ($D=3$) and a mixture with a relatively large number of denoisers ($D=7$). We remark that as we scale the number of denoisers, the total model size (total number of parameters) remains constant for fairness of comparisons. This also matches a hypothesis of an FPGA-like hardware configuration where total maximum power and logic gates are fixed, and we split the resources depending on $D$. 
\begin{table}[t]
\centering
\caption{Performance Comparison with different denoisers number in ICVL with mixture noise}
\label{tab:performance_via_number}
\begin{tabular}{cccc}
\textbf{Model} & \textbf{PSNR} & \textbf{SSIM} &\textbf{SAM} \\
\hline
\hline
{$D=1$}  & 39.67 dB & 0.9506 &0.0772 \\
{$D=3$}  & 39.45 dB & 0.9473 &0.0833 \\
{$D=5$} & 39.42 dB & 0.9468 &0.0811 \\
{$D=7$}  & 39.29 dB & 0.9470&0.0857 \\
\hline
\end{tabular}
\end{table} 

\textbf{Peak performance.}  In this experiment, we evaluate the our peak performance with different numbers of denoisers, following the same setting as Table \ref{tab:comparison}. As shown in Table~\ref{tab:performance_via_number}, increasing the number of denoisers leads to slight degradations in PSNR and SAM. This is expected due to fixed number of total parameters and the general behavior of ensembles of neural networks, where a few larger models outperform large ensembles of smaller models.

\textbf{Power Scalability.} We now investigate the power scalability behavior as a function of  number of denoisiers. For this experiment we set $\lambda=0$ and use the ICVL training set with mixture noise pattern. The results is shown in Fig.\ref{fig: diff_denoiers_power}. Similarly to the Fig.\ref{fig: power scalability}, the solid curve denotes the average performance and the error bar presents the variability. However, the x axis in Fig.\ref{fig: diff_denoiers_power} denotes the normalized number of denoisers (denoisers used divided by $D$) and can be interpreted as the percentage of maximum power actually used.
We observe that increasing the number of denoisers improves power scalability by providing finer-grained performance levels at different power budgets. However, this comes at the cost of slightly reduced image quality, consistent with the peak-performance analysis.

\textbf{Fault Resilience.} In this section, we replicate the experiment in Sec. \ref{sec:experiment_fault} to compared the resilience to faults of the $D=3,5,7$ mixture configurations. The result is shown in fig.\ref{fig:diff_denoisers_fault_injection}. We can clearly see that using a larger number of denoiser improves the fault resilience of the system by leading to superior image quality at a given fault probability.

\section{Conclusions and Limitations}
\label{sec:conclusions}

We addressed the design problem of neural networks onboard satellites by postulating three goals: i) high inference quality at low complexity; ii) power scalability; iii) fault tolerance. Based on that, we presented a mixture-of-denoisers model with a line-wise architecture for HSI denoising that meets the demands of real-time processing, while addressing the power scalability and fault tolerance goals. The main limitation of the proposed design is the causality of the line-wise processing which cannot exploit the information of future lines, limiting quality. Future work will address hybrid designs that while maintaining the line-based approach for memory efficiency may use bidirectional processing when coupled with a small buffer of lines. 

\bibliographystyle{IEEEtran}


\end{document}